\documentclass[11pt,letterpaper]{article}
\pdfoutput=1
\usepackage{graphicx,array}
\usepackage{color}
\usepackage{latexsym}
\usepackage{amsthm}
\usepackage{amsmath}

\usepackage{amssymb}
\usepackage{pdfrender}

\usepackage{dsfont}
\usepackage{epsfig}
\usepackage{slashed}
\usepackage{bbold}
\usepackage{psfrag}
\usepackage[svgnames]{xcolor}
\PassOptionsToPackage{caption=false}{subfig}
\usepackage{subcaption}
\usepackage{xfrac}
\usepackage{multirow}
\usepackage{booktabs}
\usepackage[colorlinks=true,linkcolor=MediumBlue,citecolor=Green,urlcolor=violet]{hyperref}
\usepackage{cite}
\usepackage[normalem]{ulem}

\newcommand{\be}{\begin{equation}}
\newcommand{\ee}{\end{equation}}
\newcommand{\bea}{\begin{eqnarray}}
\newcommand{\eea}{\end{eqnarray}}

\newcommand{\re}{\mathrm{Re}}
\newcommand{\im}{\mathrm{Im}}

\usepackage{mathrsfs}

\def\({\left(}
\def\){\right)}

\usepackage{soul}

\begin{document}
\begin{flushright}
\end{flushright}
\vspace{.6cm}
\begin{center}
{\LARGE \bf  An almost elementary Higgs:\\ Theory and Practice }

\bigskip\vspace{1cm}{Daniele Barducci$^a$, Stefania De Curtis$^b$, Michele Redi$^b$ and Andrea Tesi$^b$}
\\[7mm]
 {\it \small

$^a$ SISSA and INFN sezione di Trieste, Via Bonomea 265; I-34137 Trieste, Italy\\
$^b$ INFN sezione di Firenze, Via G. Sansone 1; I-59100 Sesto F.no, Italy\\
 }

\end{center}

\bigskip \bigskip \bigskip \bigskip

\centerline{\bf Abstract} 
\begin{quote}
We study models that interpolate between an elementary and a composite Higgs boson. 
Such models, arising in theories with new vector-like fermions with electro-weak quantum numbers and charged under a confining gauge interaction,
are entirely compatible with current data, with only weak bounds from flavor, CP-violation and precision tests. 
After classifying the models from the point of view of symmetries, we study their collider phenomenology at LHC. 
In the most relevant scenarios, bounds from present searches exclude heavy scalar isospin triplets and quintuplets up to $\sim 200\;$GeV and we show how dedicated searches of simple signals such as $pp\to 3 \gamma W$ could improve the reach by at least a factor of 2 with present data, reaching  $O(1\;{\rm TeV})$  with higher integrated luminosities. States that mix with the SM Higgs can be tested in a variety of final states, such as $2b2\gamma$ searches relevant for double Higgs production.
\end{quote}

\newpage
\tableofcontents
\section{Introduction}
\label{sec:intro}

The discovery of the Higgs boson with the (so far) exact properties predicted by the Standard Model (SM) and the absence of any new particle has overthrown
conventional approaches to physics beyond the SM based on the concept of naturalness of the electro-weak scale. 
While the room for ``natural'' new physics has become narrow and perilous , new physics not directly related to the hierarchy problem 
and possibly motivated by different considerations such as the existence of Dark Matter is still plausible at energies accessible at the LHC or future colliders.

In this work we overview and extend studies on strongly coupled extensions of the SM that do not break the electro-weak symmetry \cite{Kilic:2009mi}.
This scenario is realized by adding to the SM lagrangian new vectorial fermions with SM charges charged under a new gauge force that confines at a scale $m_\rho$. The spectrum in the condensed phase corresponds to several hadronic states with masses around $m_\rho$, while the presence of lighter pion-like states with SM charges depends on the presence of fermions with masses smaller  than $m_\rho$.
This very simple setup has two main positive aspects: $i)$ the gauge structure completely determines the accidental symmetries that explain the lightness of the pion states and allows to determine the phenomenology in terms of a fundamental description;
$ii)$ this scenario, despite the strong coupling and the presence of matter charged under SM interactions, is quite safe from precision constraints and therefore allows a rather low overall scale $m_\rho$,  which then calls for an immediate exploration at colliders.

We focus on fermionic matter charged only under $SU(2)_L \times U(1)_Y$ and where Yukawa couplings exist between the elementary Higgs and the new fermions. The presence of this additional portal between the elementary and composite sectors does not spoil the robustness of the gauge theory construction but rather it selects an even more specific scenario, since a second 
composite Higgs is automatically present in the spectrum \cite{Antipin:2015jia,Agugliaro:2016clv}. 
The dynamics is such that the electro-weak symmetry breaking (EWSB)  is driven by the elementary Higgs but the elementary and composite 
Higgses mix so that the observed particle interpolates between these two states.  Such theories are for the Higgs the precise analog of fermionic partial compositeness  widely discussed  literature, see \cite{Panico:2015jxa} for a review. 

While the presence of an elementary Higgs makes these theories tuned, they nicely house Dark Matter
candidates that are granted by the accidental symmetries of the theory~\cite{ACDM}. Moreover they can be used for different 
approaches to the hierarchy problem such the relaxion mechanism \cite{Graham:2015cka,Antipin:2015jia,Batell:2017kho}.

In the first part of this paper we characterize the framework from the point of view of symmetries. Partial compositeness of the Higgs is linked to the breaking of parity while CP violation has interesting effects for EDMs.  After having outlined the main constraints from precision physics  we next discuss the collider signals. The lightest states of the spectrum, on which we focus, are Nambu-Goldstone bosons
(NGBs) with electro-weak charges. Of these, singlets, triplets and quintuplets  of isospin arise from NGBs made of same species fermions while a doublet requires different species that allow for 
Yukawa couplings with the elementary Higgs. The first class of NGBs can decay through anomalies into pairs of SM gauge bosons leading striking signatures such as 4 electro-weak gauge boson final states. We revisit current 
LHC searches showing that in the most favourable scenario their reach could test triplet and fiveplet masses up to 500 GeV with current data with small variations of current analyses and that the reach could be greatly improved with slight changes of the experimental strategies.
We also consider different signals such as heavy Higgs production. While some features are common with type-I Two Higgs doublet models and the heavy Higgs can be tested in a variety of final states, its composite nature allows for abundant exotic decays into a singlet and a SM Higgs giving rise to $bb\gamma\gamma$ final states that are strongly constrained by double Higgs searches. 

\section{Electro-weak preserving strong dynamics}

The class of models we consider is defined by the presence of new fermions 
vectorial under the SM gauge group and charged under a new (dark) gauge group that confines around the TeV scale.
They are described by the renormalizable lagrangian 
\be\label{definition}
\mathscr{L}=\mathscr{L}_{\rm SM} + \mathscr{L}_{\rm VLC}+\mathscr{L}_{\rm mix}
\ee
where $\mathscr{L}_{\rm VLC}$ contains only the kinetic terms of the new fermions.
Differently from composite Higgs models with fermion partial compositeness, here the flavor structure is identical to the SM one, 
and the mixing lagrangian only contains renormalizable interactions between the SM fields and the vector-like fermions. Therefore, the only allowed interactions are 
gauge and Yukawa interactions. The dynamics of the theory is such that strong dynamics confines without breaking the electro-weak symmetry, which is an effect induced by the elementary Higgs.

We will focus on models with constituents in the fundamental representation of  $SU(N)$, $SO(N)$ or $Sp(N)$ gauge and only electro-weak charges 
that are most relevant for partially composite Higgs models  \cite{Antipin:2015jia}.
Motivated by unification (and simplicity) we consider different combinations of\footnote{Other  representations compatible with $SU(5)$ unification are $E=(n,1,-1)$ and $T=(n,3,1)$.},
$$
N = (n,1)_0\,,~~~~~~~~~~~~L=(n,2)_{-\frac12}\,,~~~~~~~~~~~~~~V=(n,3)_0
$$
with the same quantum numbers under the SM as Bino, Higgsino and Wino in supersymmetry. 
In order for the matter content to be anomaly free it is enough to have real representations $R=\psi+\psi^c$
where $\psi^{(c)}$ are Weyl spinors with conjugate charges.

Each field can have a vectorial mass and the following Yukawa couplings are allowed,
\begin{equation}\label{mixing}
\mathscr{L}_{\rm mix}= y_N H L N^c+ \tilde{y}_N H^\dagger L^c N + y_V H L V^c+ \tilde{y}_V H^\dagger L^c V + m_V V V^c + m_L L L^c + m_N N N^c + h.c.
\end{equation}
Such lagrangian contains in general two perturbative CP violating phases corresponding to the relative phase of $y$ and $\tilde y$. The mass matrix is
analogous to the one of neutralinos in supersymmetry with the notable difference that the fermions are also charged under the dark force.
Notice that this expression is valid in general, although  for $SO(N)$ and $Sp(N)$  $V^c=V$ and $N=N^c$ are Majorana fermions. 
Upon confinement bound states are formed. In this paper we will assume that $m_Q< \Lambda_{\rm DC}$ so that the lightest states are scalar pions
and spin-1 resonances (for the opposite regime see \cite{Mitridate:2017oky,Elor:2018xku}). Their quantum numbers are fixed by the symmetries.
At the constituent level the pions correspond to the fermion bi-linears
\begin{eqnarray}
V \times V^c& =& \eta+ \pi_a+\phi_{ab} \nonumber \\ 
L\times L^c& =&  \eta+ \pi_a \nonumber \\ 
L\times N^c&=& K_\alpha \nonumber \\
L\times V^c&=& K_\alpha + H_{a\alpha}
 \end{eqnarray}
where $\eta$ is an $SU(2)_L$ singlet, $\pi_a$ a triplet, $\phi_{ab}$ a quintuplet (described by a symmetric traceless 3 by 3 matrix), $K_\alpha$ a doublet and $H_{\alpha a}$ a quadruplet.
In ${\rm SO}(N)$ and ${\rm Sp}(N)$ theories the reality of the reps eliminates some  of the pions. For example $V$ in ${\rm SO}(N)$ gives rise to an isospin quintuplet
while in ${\rm SU}(N)$ also a triplet exists.

\subsection{Dynamics in the condensed phase}
Below the confinement scale the low energy physics is described by a QCD-like chiral lagrangian. 
For $SU(N)$ and $SO(N)$ gauge theories with fundamental fermions (generalization to $Sp(N)$ is also possible)  the condensate that breaks spontaneously the global symmetries is given by
\be
\langle \psi_i \psi_j^c \rangle=-g_\rho f^3 \delta_{ij}
\ee
where $g_\rho \approx 4\pi/\sqrt{N}$. The associated Goldstone bosons are described by the unitary matrix $U=e^{i\sqrt{2}\Pi/f}$ of the broken generators.
At $O(p^2)$  the low energy effective lagrangian has the general form, 
\begin{eqnarray}
\mathcal{L}&=&\frac{f^2}{4}{\rm Tr} [D_\mu U D^\mu U^\dagger]+  g_\rho f^3 {\rm Tr}[MU^\dagger+h.c.] + \frac {3 g_2^2 g_\rho^2 f^4 } {2(4\pi)^2} \sum_{i=1..3} {\rm Tr}[U T^i U^\dagger T^i]
\label{lagrangianN}
\end{eqnarray}
where $M$ is the quark mass matrix originating from Eq.~\eqref{mixing} that, by construction, is linear in the Higgs field. Expanding $U$ around the origin one finds\footnote{In the presence of a non zero
$\theta$-angle in the dark sector these formulas continue to apply by judiciously rotating the phases such that no linear coupling of the singlet appears, see Appendix~\ref{appendixA}.},
\begin{equation}
{\cal L} \subset -m_K^2 |K|^2 - i y_- g_\rho f^2  (b  K^\dagger H+ h.c.) +  y_+ g_\rho f \left(a_1 \eta K^\dagger H + a_3 \pi^a K^\dagger \sigma^a H + h. c.\right)
\label{mixing-trilinear}
\end{equation}
where $m_K^2\sim (m_{Q_1}+m_{Q_2}) g_\rho f+ \Delta^{gauge}$ depends on the vectorial fermion masses and gauge contribution and we do not explicitly write interactions with quadruplets and quintuplets.The coefficients $a,b$ are $O(1)$ that depend on the specific model ($a_1=1/\sqrt{6}$, $a_3=-1/\sqrt{2}$ and $b=\sqrt 2$ for the $L+N$ model, see Appendix \ref{appendixA}) and we defined the combinations
\begin{equation}
y_-\equiv(y-{\tilde y}^*)~~~~~~~~~~~~{\rm and}~~~~~~~~~~~~~y_+\equiv(y+{\tilde y}^*)\,.
\end{equation}
The lagrangian above encodes the mixing between elementary Higgs $H$ and composite Higgs $K$. 
The mixing angle  reads
\begin{equation}
\epsilon\equiv i b \frac {y_-}{g_\rho} \frac{m_\rho^2}{m_K^2},
\label{eq:mixing}
\end{equation}
where we introduced the mass scale of the theory $m_\rho\sim g_\rho f$. We will mainly work in the mostly-elementary regime where $\epsilon\ll 1$, which justifies our approximations,
see \cite{Agugliaro:2016clv} for the opposite regime.

Pions made of identical species decay to two SM gauge bosons through anomalies,
\begin{eqnarray}
\mathscr{L}_{F\tilde{F}}&=& -\frac{1}{16\pi^2} \frac{\eta}{f} \left(g_1^2 c^{\eta}_{BB} B_{\mu\nu} \tilde B^{\mu\nu}+g_2^2 c^{\eta}_{WW} W_{\mu\nu}^a \tilde W^{a \mu\nu}\right)\nonumber \\
&&- c^{\pi}_{WB} \frac{g_1 g_2}{16\pi^2} \frac{\pi_a}{f}  W_{\mu\nu}^a \tilde B^{\mu\nu}- c^{\phi}_{WW} \frac{g_2^2}{16\pi^2}\frac{\phi_{ab}}{f} W_{\mu\nu}^a\tilde{W}^{b\,\mu\nu},
\label{anomalies}
\end{eqnarray}
giving
\begin{equation}
\Gamma(\Pi \to V V)= c_{\Pi}^2  \frac{\alpha_i \alpha_j}{64 \pi^3}  \frac {m_\Pi^3}{f^2},
\end{equation}
where the coefficients $c_\Pi^2$ are indicated in Tab.~\ref{tab:anomalies}.

\begin{table}[t!]
 \begin{center}
    \begin{tabular}{ c | c | c || c || c}
  \hline 
    Model 			& $c^{\eta}_{BB}/N$   & $c^{\eta}_{WW}/N$ & $c^{\pi}_{WB}/N$ & $c^{\phi}_{WW}/N$ \\     
    \hline
      \hline
  $L+N$ 	 & 	$\frac{1}{2\sqrt{3}}$	& $\frac{1}{2\sqrt{3}}$	  & $-\frac{1}{2}$	 	&   $\slash$ \\
    \hline
  $L+V$     	 & 	$\frac 1 2 \sqrt{\frac 3 {5}}$	& 	$-\frac 1 2\sqrt{\frac 5 {3}}$  & 	 $-\frac 1 2$	&  $\frac 2 {\sqrt{3}}$  \\
       	 \hline
  \end{tabular}
\end{center}
\caption{\em Coefficients of the anomaly terms for the models with $L+N$ and $L+V$ fermion constituents.}
\label{tab:anomalies}
\end{table}

The singlet and triplets can also couple to the two Higgs bosons. From the trilinear vertices $\eta(\pi) HK$ one obtains, after diagonalization of the mass matrix at leading order in $\epsilon$, the interaction
\begin{equation}\label{HH}
\mathscr{L}_{HH}= \epsilon y_+ m_\rho   \left(a_1 \eta H^\dagger H + a_3  \pi^a H^\dagger \sigma^a H\right) + h.c.
\end{equation}
allowing for a tree level decay into two Higgs bosons.  A similar coupling is induced by the $\theta_{\rm DC}$ angle, proportional to $\epsilon^2$. Note that the same couplings are also generated through higher order terms in the chiral lagrangian, without the enhancement  $m_\rho^2/m_K^2$ contained in $\epsilon$.
After electro-weak symmetry breaking eq.  (\ref{HH}) induces a mixing between with the Higgs,
\begin{equation}
\Delta \sim \epsilon \frac{y_+}{g_{\rho}} \frac {m_\rho^2}{{\rm Max}[m_h^2, m_{\eta,\pi}^2]}  \frac v f\,.
\label{mixingeta}
\end{equation}

Pions made of different species are stable from the point of view of the strong sector. They can decay through the coupling to the elementary Higgs that
explicitly breaks species number.

\subsection{Symmetries of the model}
\label{sec:symmetries}
Since the model is completely determined by the fundamental interactions of Eq.~\eqref{mixing}, it is possible to analyze the symmetries in terms of the UV lagrangian. In general it
contains a physical phase for each pair of Yukawa couplings and the dark color $\theta-$angle.
For simplicity we discuss them in the basis where all the fermion mass terms are real: in this case possible complex phases are only in the Yukawa terms and the $\theta$-angle. A summary of the discussion of this section is provided in Tab.~\ref{simmetrie}.

\paragraph{Parity:}
The new sector is vector-like under the SM, therefore parity P is usually a good symmetry of the gauge sector.
Clearly it can be broken by the interplay between the fermion masses and the dark $\theta$-angle, as well as by the Yukawa couplings to the (elementary) Higgs field.
At fundamental level, P exchanges $\psi \leftrightarrow \bar\psi^c$ and $\vec{x}\to -\vec{x}$.

From the interactions in Eq.~\eqref{mixing}, it follows that in all the models parity is respected by requiring
\be\label{P}
\mathrm{P}: \quad\quad y_-=0,\quad \theta=0, \quad ( \mathrm{Im}[m_j]=0)
,\ee
where the indices $i$ and $j$ refer to all the monomial terms in the lagrangian. It is important to notice that some of the above requirements are redundant, since in a given model only few phases are physical.

Because the NGB are odd under spacetime parity  a mixing with the elementary Higgs is only allowed if $\mathrm{P}$ is explicitly broken.
To discuss the confined phase it is convenient to rotate $\theta_{\rm DC} G \tilde{G}$ to the mass matrix through a chiral rotation of the light quarks.
This can be done while ensuring that the pions do not acquire a VEV in the vacuum. In this basis the mixing is given by Eq.~\eqref{eq:mixing}.
We note that this mixing could vanish or be suppressed through a cancellation of the two physical phases even though $\mathrm{P}$ is broken.

\paragraph{CP:}
The new strongly interacting sector has in general a few complex phases that cannot be eliminated, thus signalling the explicit breaking of CP.  
Again, in the basis where fermion masses are real, the condition for CP conservation can be written as
\be\label{CP}
\mathrm{CP}: \quad\quad \mathrm{Im}[y_i \tilde y_i]=0,\quad \theta=0, \quad ( \mathrm{Im}[m_j]=0),
\ee
therefore in this case, by accident, when P is conserved so is CP.

Let us note that in general CP can be broken  by the Yukawa couplings or by $\theta_{DC}$.

\paragraph{Custodial symmetry:}
Custodial symmetry SO(4) is an approximate accidental symmetry of the Higgs sector in the SM, which is broken to SO(3) by the Higgs condensate. In this framework we expect additional source of custodial breaking, given that the Higgs is linearly coupled to the new sector. Notice that it is not possible to realize accidentally custodial at the renormalizable level, without assuming an SU(2)$_R$ exact symmetry. However, we can distinguish two qualitatively different behaviours depending on whether the strongly interacting sector allows for an unbroken SO(4) in the vacuum. Therefore, we can distinguish two classes of models
\begin{itemize}
\item $\mathcal{H}\supseteq SO(4)$. In this case the sigma model respects an SO(4) symmetry, and so does the condensate $\psi_i\psi^c_j$. The only possible custodial breaking, which comes entirely from the new sector, can then only be ascribed to the Yukawa couplings.
\item $\mathcal{H}\nsupseteq SO(4)$. In this case custodial symmetry is generically broken, and it can arise only accidentally because of discrete symmetries.
\end{itemize}

When more than one composite Higgs doublet exists, custodial symmetry is not in general sufficient to protect the $\hat T$ parameter because the two VEVs can have different phases \cite{Mrazek:2011iu}. Therefore these models generically behave as the ones without custodial symmetry. By imposing that the Yukawa couplings respect $SO(4)$ it is however possible to align the two VEVs so that no tree level contributions are generated.

\begin{table}[t]\small
$$\begin{array}{c|c|c|c|c|c}
\hbox{}& \hbox{fermions}& \mathcal{H} & \hbox{P} & \hbox{CP} & \hbox{SO(4)$_c$} \\
\hline
 \hbox{SU(N)} & L + N & \hbox{SU(3)} & y_-=0, \theta=0 & \mathrm{Im}[y\tilde y]=0, \theta=0 & \hbox{accidental at tree-level, if P and CP }\\
 \hbox{SU(N)}& L + V &  \hbox{SU(5)} & y_-=0, \theta=0 & \mathrm{Im}[y\tilde y]=0, \theta=0 & \hbox{accidental at tree-level, if P and CP }\\
\hline
 \hbox{SO(N)}& L + L^c + N &  \hbox{SO(5)} & y_-=0, \theta=0 & \mathrm{Im}[y\tilde y]=0, \theta=0 & y=\tilde y\\
\hline
 \hbox{Sp(N)}& L_0 + E_{\pm1/2} &  \hbox{Sp(4)}& y_-=0, \theta=0 & \mathrm{Im}[y\tilde y]=0, \theta=0 &  y=\tilde y\\
\end{array}$$
\caption{\em Symmetries of the models depending on the Yukawa couplings and $\theta-$term.}
\label{simmetrie}
\end{table}

\paragraph{Accidental symmetries of the $O(p^2)$ chiral expansion}

The leading order low energy effective action may enjoy accidental symmetries that are not true symmetries of the UV lagrangian. In an expansion up to terms of $O(m,g^2)$, the scalar potential describing the interactions of the elementary $H$ and the composite NGBs is given by $\propto \mathrm{Tr}[M U^\dagger + h.c.]$. When $M$ is hermitian, the scalar potential develops an accidental symmetry
\be\label{pion-number}
(-1)^{N_\Pi}:\quad    U \leftrightarrow U^\dag
\ee
under which $\Pi\to -\Pi$. This condition is realized accidentally in the scalar potential when CP (and P) is conserved. Clearly Eq.~\eqref{pion-number} does not represent a good symmetry of the theory as it is broken by anomalies (at the fundamental level it corresponds to $\psi \leftrightarrow \bar\psi^c$ without spatial inversions).

Another accidental symmetry that can be realized in the condensed phase at $O(p^2)$
\be
\mathrm{P}': y_+=0,\,\quad \theta=0\,.
\ee
This is a parity under which the pion doublet that mixes with the Higgs transforms as a scalar: this forbids interactions among the pion doublet, $H$ and an odd number of other pions. Particularly, if $P'$ is broken we expect decay modes $K\to H + \pi$.

\section{Precision constraints on the almost elementary Higgs}

The presence of a strong sector that confines at a scale $m_\rho$ close to the TeV can manifest itself in a variety of experimental probes. Constraints from precision observables would depend upon the way the Higgs, the electro-weak and the flavor sectors are coupled to the strong dynamics. Hence, already from the previous discussion, we expect all the effects to be screened by \textit{at least} two powers of $y/g_\rho$ or $g/g_\rho$, where $y$ and $g$ indicate SM Yukawa and gauge couplings, depending on which precision observable we are considering, see also \cite{Antipin:2015jia}.

\subsection{Higgs couplings}
Modifications to Higgs couplings are a robust predictions of this class of models and they originate because of two main effects: the existence of two Higgs doublets and the fact that the second doublet $K$ is composite.
In order to properly discuss the physics of the Higgs sector it is important to notice that, upon expansion of Eq.~\eqref{lagrangianN}, the relevant interactions affecting the SM Higgs couplings are the one involving the pion doublet $K$,
\be\label{lag-coupling}
\mathscr{L}\supset |D_\mu H|^2 + |D_\mu K|^2 + \frac{c_K}{2f^2}(\partial_\mu |K|^2)^2 - \epsilon m_K^2( K^\dag H + h.c.)+ y_u \bar{Q}_L  \tilde H u_R + y_d \bar{Q}_L  H d_R+ y_e \bar{L}_L  H e_R\,.
\ee
We would like to stress the two different sources of possible contributions to the modified Higgs couplings. 
After integrating out the $K$ doublet from the above lagrangian the third term, that originates from the non-linearities of the $\sigma-$model, gives rise to the operator
\be
\frac{c_K |\epsilon|^4}{2f^2} O_H
\ee
where in the notation of~\cite{Giudice:2007fh}, $O_H=(\partial_\mu |H|^2)^2$. This term gives an universal relative shift to Higgs couplings due to partial compositeness
\be
\frac{g_{h}}{g^{SM}_h}\bigg|_{\rm comp.}=1- c_K |\epsilon|^4\frac{v^2}{f^2},
\ee
up to corrections of order $v^4/f^4$.
On the other side the term in $V(H,K)$ that mixes the elementary Higgs and the composite K, see Eq.~\eqref{mixing-trilinear}, will contribute to Higgs coupling modifications that are common to a type-I 2HDM. In this case it is known that the modifications are of the form \cite{Branco:2011iw}
\be
\frac{ g_{hff}}{g^{SM}_{hff}}\bigg|_{\rm type-I}=\cos\delta -\frac{\sin\delta}{\tan\beta},\quad\quad \frac{ g_{hVV}}{g^{SM}_{hVV}}\bigg|_{\rm type-I}=\cos\delta
\ee
where $\delta$ is the rotation from the \emph{Higgs basis} (the basis where only the SM like Higgs gets a VEV) and $\beta$ is the rotation from Eq.~\eqref{lag-coupling} to that basis. Notice that given the peculiar linear interaction of $H$ with the composite sector (that we assume will not break the electro-weak symmetry unless coupled to the elementary sector), $\langle K \rangle/\langle H\rangle = 1/\tan\beta = \epsilon$. By performing the diagonalization of the $H$ and $K$ mass lagrangian one gets
\be
\delta\approx -|\epsilon|\frac{m_h^2}{m_K^2}.
\ee
We therefore  expect  the leading corrections to the Higgs couplings to be the ones due to the 2HDM structure rather than the ones arising from the  partially composite nature of the Higgs boson. In an expansion in $\epsilon$ and $m_h^2/m_K^2$ one should expect the following size of deviations
\be
\frac{\delta g_{hff}}{g_{hff}}=  |\epsilon|^2 \frac{m_h^2}{m_K^2} -c_K |\epsilon|^4\frac{v^2}{f^2},\quad\quad \frac{\delta g_{hVV}}{g_{hVV}}= - |\epsilon|^2 \frac{m_h^4}{2m_K^4} -c_K |\epsilon|^4\frac{v^2}{f^2},
\ee
where in the $L+N$ model $c_K=1$.

\subsection{Electro-weak precision tests}

Despite the almost elementary nature of the Higgs bosons, we do expect several contributions to the electro-weak precision observables. Differently from effects in the Higgs couplings, here the compositeness of $K$ plays an important role. This is due to the fact that electro-weak tests are sensitive to operators with $H$ and transverse gauge bosons and usually requires less insertions of $\epsilon$.
While in general some contributions are not calculable in a given QCD-like theory, we can estimate the size of the coefficients based on the structure of the operators listed in Tab.~\ref{tab:ops}. In particular, for the $T$ parameter, given the non-linearities of the chiral lagrangian of Eq.~\eqref{lagrangianN} we have an operator of the form
\be
\frac{1}{f^2} (K^\dag \overleftrightarrow D_\mu K)^2, 
\ee
which breaks the SO(4) symmetry acting on the real components of $K$. Therefore by four insertions of $\epsilon$ we generate the operator contributing to the $T$ parameter,
\be
\hat{T}\sim \epsilon^4 \frac {v^2}{f^2}
\ee
On the other side the $S$ parameter is due to the tree-level exchange of $SU(2)_L$ and $U(1)_Y$ spin-1 resonances, that can be estimated as
\be
\hat{S}\sim \epsilon^2\frac{m_W^2}{m_\rho^2}
\ee
All these effects allow for a confinement scale below the electro-weak scale for a moderate mixing parameter. 
Since the contribution to electro-weak symmetry breaking is suppressed, the only model independent effect is provided  by oblique corrections at $O(p^4)$.  The deviations from the SM can be 
parametrized in terms of the $X, Y, W, Z$ parameters defined in~\cite{Barbieri:2004qk} from the two point functions of gauge fields or equivalently in terms of four-fermi operators. For example, normalizing the operators as in~\cite{Farina:2016rws} we find
\begin{equation}
W \sim \frac{m_W^2}{m_\rho^2} \frac{g_2^2}{g_\rho^2}= N \frac {\alpha_2}{4\pi} \frac{m_W^2}{m_\rho^2} \sim 5\times 10^{-5} \left(\frac {N}3\right) \left(\frac {\rm TeV}{m_\rho}\right)^2
\end{equation}
A similar contribution is generated for $X$ and $Y$ if the fermions have hypercharge. With our assumptions instead $Z=0$ since no coloured  states are included.  

\begin{table}[h!]\small
$$\begin{array}{c|c|c|c|c}
\multirow{ 2}{*}{Coeff} & \multirow{ 2}{*}{Operator} & \multirow{ 2}{*}{\hbox{Present bound }$[\times 10^{-3}$] } & \hbox{LHC13 bound } [\times 10^{-3}]   \\
&  &  & 300\;{\rm fb}^{-1} \quad \;\;\;3000\;{\rm fb}^{-1}   \\ 
\hline\hline
\hat S  & v^{-2} s_W c_W^{-1} (H^\dag W_{\mu\nu} H) B^{\mu\nu}  & \hat{S}=0.86\;\;[\hat T = 0] & $\slash$  \;\;\;\;\quad \quad\quad \quad $\slash$ \\
\hat T  &  1/2 v^{-2} (H^\dag \overleftrightarrow{D_\mu} H)^2 &\hat{T}=1.19\;\;[\hat S=0] & $\slash$  \;\;\;\;\quad \quad\quad \quad $\slash$ \\
W  &  -2 v^{-2}J_{L,\mu}^a J_{L,a}^\mu & W=0.3 & 0.07 \;\;\quad \quad \quad 0.045 \\
Y &  -2 v^{-2}t^2_WJ_{B,\mu} J_{B}^\mu& Y=0.4 & 0.23 \;\;\;\;\quad \quad \quad 0.12 \\
\end{array}$$
\caption{\em \label{tab:ops} Operators relevant for the electro-weak fit. The bound on $\hat S$ and $\hat T$ are taken from~\cite{Baak:2012kk}. The present bound on $W$ and $Y$ is due to LEP. Future bounds assume the projection of~\cite{Farina:2016rws}.}
\end{table}

\subsection{Flavor and CP bounds}
The only sources of breaking of the flavor symmetries are the Yukawa couplings so that our theories automatically realize Minimal Flavor Violation (MFV) (see \cite{MFV} for a detailed discussion).
The main effects from new physics arise due to the exchange of the charged Higgs bosons $K^\pm$ (and smaller effects from the charged component of the triplet)
which contributes at tree-level to charged current $\Delta F=1$ processes (such as $B\to \tau \nu$ and $K\to \mu \nu$) and at one-loop to neutral current $\Delta F=1,2$ transitions (such as $b\to s \gamma$). However, the new effects are aligned in flavor space because of the MFV set up, therefore they manifest themselves mainly just as an overall rescale of the SM rates. The bounds on the parameters of the model are then only sensitive to the precision of the determination of the SM effects, which have at best a few percent accuracy.

The loop induced corrections are mainly encoded in the Wilson coefficients to dipole operators $\mathcal{O}_{7,8}$ (see \cite{MFV} for the notation) contributing to $b\to s \gamma$ transitions,
\be
\mathscr{L}\supset \frac{4G_F}{\sqrt{2}}V_{tb}V_{ts}^* \frac{e m_b}{16\pi^2} \bigg( C_7 \bar s_L \sigma^{\mu\nu} b_R F^{\mu\nu} + C_8 \bar s_L \sigma^{\mu\nu} T^a b_R G^{a}_{\mu\nu} \bigg) + h.c.
\ee
In our model the new physics contribution arises because of the exchange of $K^\pm$ in the loop (see \cite{Enomoto:2015wbn} for a study in 2HDM), therefore it is suppressed by two insertions of the $K$ coupling to fermions and by the mass of $K$. The predictions for the Wilson coefficients are
\be
C_7^{NP}\approx \epsilon^2\, \frac{m_t^2}{m_K^2} \frac{1}{3}\bigg( \log \frac{m_t^2}{m_K^2} -\frac{25}{24}\bigg)\, ,\quad C_8^{NP}\approx \epsilon^2\, \frac{m_t^2}{m_K^2} \frac{1}{2}\bigg( \log \frac{m_t^2}{m_K^2} -\frac{5}{3}\bigg)
\ee
The deviation from the SM is constrained within a 10\%.

Another contribution is the modification of $B\to \mu \mu$. In this scenario the largest contribution arises from penguin diagrams with a charged $K$ in the loop, which give
\be
\frac{\delta \mathcal{B}(B\to \mu\mu)}{ \mathcal{B}(B\to \mu\mu)}\approx 2 \frac{C_{10}^{NP}}{C_{10}^{SM}}\approx \frac{1}{4 C_{10}^{SM}}\, \epsilon^2 \ \frac{m_t^2}{m_K^2} (\log \frac{m_t^2}{m_K^2} + 1)
\ee
Other loop induced processes, such as $\Delta B=2$, are very well measured experimentally (per mille accuracy), but have to face large theoretical uncertainties, which limit their capabilities to constrain type-I two Higgs doublet models (see also \cite{Meroni}). All these bounds are less constraining than the ones from precision measurements in the electro-weak and Higgs sector.

On the other hand for complex Yukawa couplings the theory contains extra CP violating phases that
induce Electric Dipole Moments for SM particles. For the electron one finds \cite{Agugliaro:2016clv}
\begin{equation}
d_e \sim  10^{-26}\, {\rm e\,cm} \times  {\rm Im}[y_- y_+^*]   \times \left(\frac {\rm TeV}{{\rm Min}[m_{\pi_{3},\eta}]}\right)^4 \times \left( \frac {m_\rho}{\rm TeV}\right)^2
\end{equation}
to be compared with the experimental limit $d_e < 8.7 \times 10^{-29}\, {\rm e\,cm}$ at $90\%$ C.L. \cite{Baron:2013eja}.

\subsection{Summary of precision constraints}

\begin{figure}[t!]
\begin{center}
\includegraphics[width=0.45\textwidth]{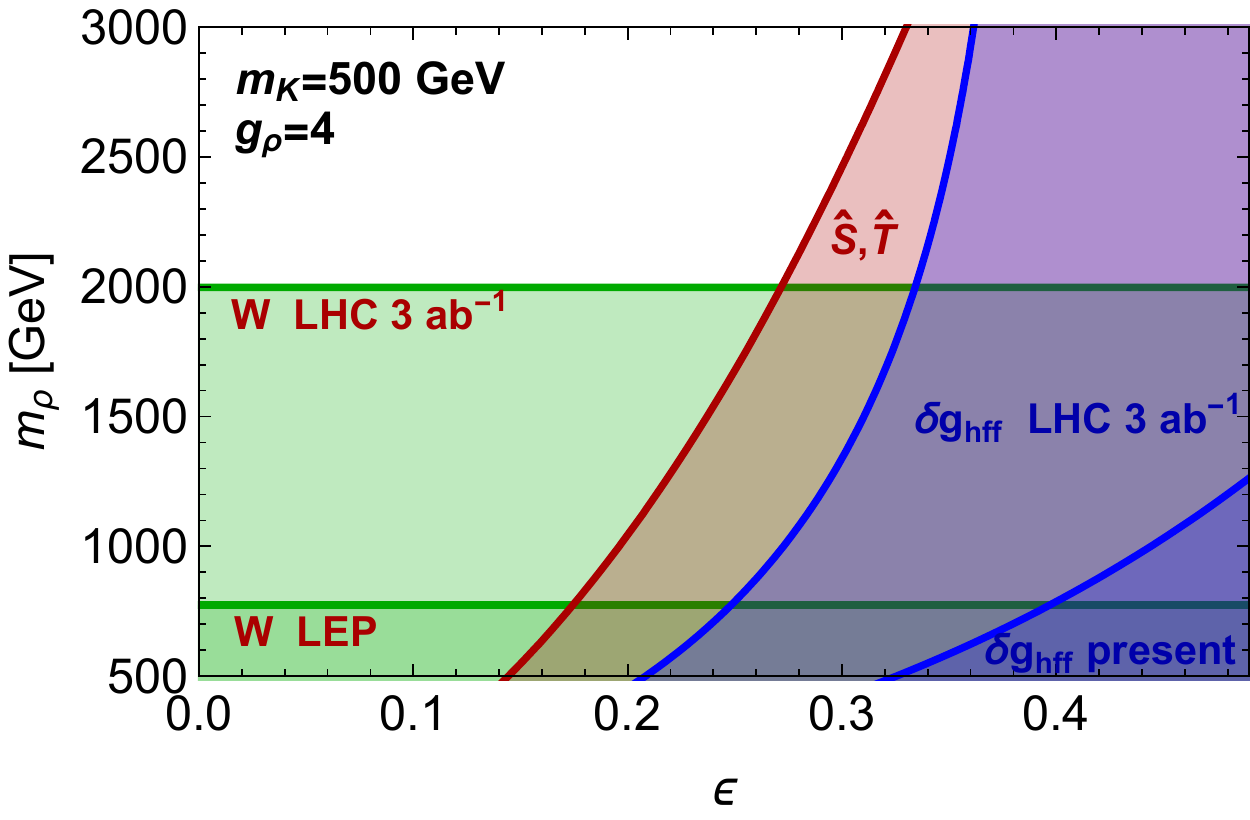}\quad
\includegraphics[width=0.45\textwidth]{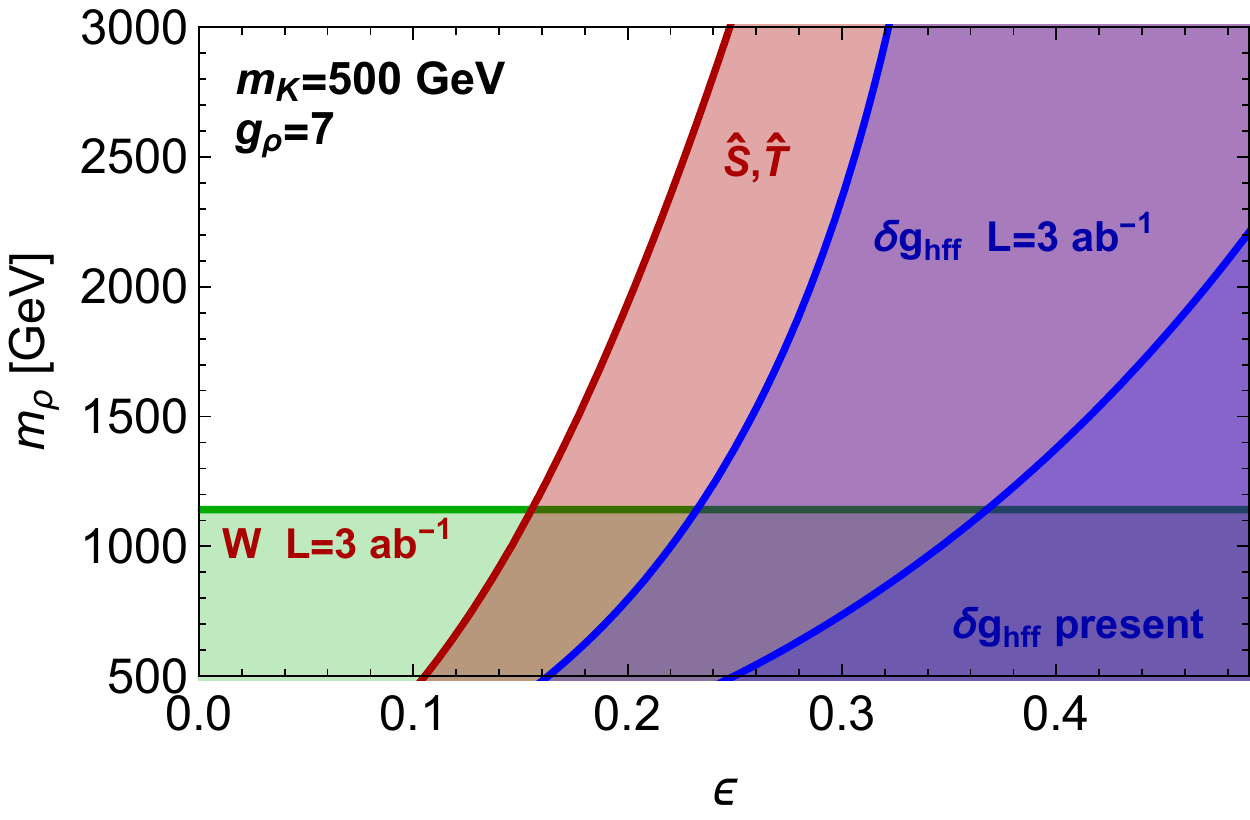}
\caption{\label{fig:indirect} \em Summary of indirect constraints in the $\epsilon$-$m_\rho$ plane for $g_\rho=4$ (left) and $g_\rho=7$ (right). The bounds on the Higgs coupling assume a 5\% deviation for present data and 1\% deviation for future data.}
\end{center}
\end{figure}

The bounds from precision constraints arising from Higgs coupling measurements and electro-weak precision observables
are summarized in Fig.~\ref{fig:indirect} projected in the $\epsilon$-$m_\rho$ plane while, as mentioned, flavor bounds turn out to be less constraining.
We see that the current strongest bounds still arise from the measurements of the $\hat S$ and $\hat T$ parameter performed at LEP and that they will not be surpassed by future precision test of the Higgs couplings, by assuming an uncertainty on their determination of  $\sim\;1$\%.
On the other hand current bounds on the $W$ parameter are able to set a bound of $m_\rho\sim 700\;{\rm GeV}$ only for moderate value of $g_\rho$ while the  measurement of the transverse mass spectrum of charged current Drell-Yan, as proposed in~\cite{Farina:2016rws}, will be able to enforce a bound on $m_\rho$ of 2 TeV and 1 TeV with  $g_\rho=4$ and $g_\rho=7$ respectively.

\section{LHC collider limits}
\label{collider}

In this Section we outline the main strategies to search for models with an almost elementary Higgs at the LHC, see~\cite{Kilic:2010et} for previous related studies. 
The signatures are production of spin-0 NGBs and heavier spin-1 resonances with electro-weak quantum numbers.
For concreteness we focus on $SU(N)$ models with fermions $L+N$ where the lightest particles are a singlet $\eta$, an electro-weak doublet $K$ and a triplet $\pi$.
The  $O(p^2)$ lagrangian is given explicitly in appendix~\ref{appendixA}. For our analysis we will also include the lightest spin-1 resonances, an $SU(2)_L$ triplet with couplings
\begin{equation}
\label{pheno}
\mathscr{L}_{INT}=  g_\rho \rho_\mu^a \left(K^\dag \frac{\sigma^a}2 \overleftrightarrow D_\mu K+\pi^T T^a \overleftrightarrow D_\mu \pi \right) +\frac{g^2}{g_\rho} \rho_\mu^a \left(i H^\dag \frac {\sigma^a}2 \overleftrightarrow D_\mu H  + \bar{f}_L \frac {\sigma^a}2 \gamma^\mu f_L\right)\,.
\end{equation}

As emphasized above the theory is very constrained by the symmetries of the fundamental lagrangian. In the following Sections we discuss the main phenomenology based on the size of  $y_\pm$, and since the qualitative behavior strongly depends  on their relative size we find convenient to analyse the asymptotic cases in order to simplify the discussion. The real situation can be often an interpolation of the extreme cases. A rough schematic summary of the outcome of the phenomenological analysis can be found in Tab.~\ref{tab:scenario}.
\begin{table}[h]
\begin{center}
\begin{tabular}{c|c|c}
\multicolumn{3}{c}{$\eta$ singlet, $\pi$ triplet}\\  \hline\hline
 Scenario & Production & Decay \\
 \hline
  $y_- y_+ \ll 1$ & EW $(\pi \pi)$
 & $\eta, \pi \to V_T V_T$ \\
  $y_- y_+ \lesssim 1$ & $gg\to \eta, \pi^0$ & $ \eta, \pi^0 \to V_L V_L$, $t\bar t$ \\
\end{tabular}
\quad\quad
\begin{tabular}{c|c|c}
\multicolumn{3}{c}{$K$ doublet}\\ \hline\hline
 Scenario & Production & Decay \\
 \hline
  $y_- \ll y_+$ & $gg\to K$ & $ K \to H \eta$ \\
  $y_- \gg y_+$ &  $gg\to K$  &  $K\to t\bar t, V_L V_L$ \\
\end{tabular}
\caption{\em \label{tab:scenario} Summary of different regimes for the pion production and decays.}
\end{center}
\end{table}\\
The pions, except for the singlet $\eta$, can be pair produced  via electro-weak interactions. 
Single production through gluon fusion is present only when a sizeable coupling to the SM top quark is present while single production through electro-weak gauge bosons is negligible. Depending on the quantum number of the NGB, these additional (non-universal) couplings are generated before and/or after electro-weak symmetry breaking, thus substantially changing  the quantitative analysis.
In the following for each resonance we will discuss its main production and decay modes. 
When available we will recast existing limits from LHC searches and in some cases we will comment on how to improve the reach of present analyses.

\subsection{$\eta$ singlet}
We start our phenomenological discussion with the singlet $\eta$, which is a rather elusive particle among all the pNGBs.
It cannot be produced substantially through the electro-weak anomalies in the kinematic region that we are interested in, whereas it can be sizeably produced in the decay of other particles (for example decay $K\to H\eta$ that we will discuss next) or after electro-weak symmetry breaking when it acquires a coupling to fermions. From the mixing (\ref{mixingeta}) a coupling to gluons is generated via a top-loop.
Gluon fusion is than the dominant production channel,
\begin{equation}\label{eq:gg-eta0}
\Gamma(\eta \to gg)\approx \epsilon^2 \frac{|y_+|^2}{g_\rho^2}\frac{v^2}{f^2}\frac{m_\rho^4}{m_\eta^4} \Gamma(H\rightarrow gg)|_{m_H=m_\eta},
\end{equation}
 At high masses (above dibosons and $t\bar t$) the phenomenology of the singlet is similar to the one of the triplet that we will analyze in Sec.~\ref{tree-level-regime}. However, the singlet, much differently from the rest of the pions, can also be quite light and so have sizeable decays to diphoton (and/or $Z\gamma$) as well as to $b\bar b/\tau\tau$ and $gg$. Moreover, given that when it is sufficiently light it can appear in decay chains of composite pions, it is very important to determine which is the dominant decay mode for a light $\eta$. There are basically two scenarios determined by the following ratio
\be\label{ratio-eta}
\frac{\Gamma(\eta\to \gamma\gamma)}{\Gamma(\eta\to b \bar{b})}\sim \frac{\alpha^2}{8\pi^2}\frac{1}{\epsilon^2}\frac{g_\rho^2}{y_{+}^2}\frac{m_\eta^2}{m_b^2}\frac{m_\eta^4}{m_\rho^4}
\sim\left(\frac {0.05}{\epsilon}\right)^2\left(\frac {0.05}{y_+}\right)^2 \left(\frac {\rm TeV}{m_\rho}\right)^4 \left(\frac {m_\eta}{100\,{\rm GeV}}\right)^6
\ee

\begin{itemize}
\item $\eta\to \gamma\gamma$: This decay channel is the dominant one for $m_\eta\approx 100 \mathrm{GeV}$, and $\epsilon \lesssim 0.05$ and $y_{+}\lesssim 0.05$. However in this regime, single production via gluon fusion is suppressed by a factor of $\lesssim 10^{-3}$ with respect to the corresponding SM Higgs cross-section, so that it is not constrained by diphoton searches (see however \cite{Mariotti:2017vtv} for a detailed description). 
\item $\eta\to b \bar{b}$: The singlet will be hidden under a huge QCD background.
\end{itemize}

\subsection{$\pi$ triplet}

The triplet decays thorough anomalies to electro-weak gauge bosons. For mass larger than 200 GeV it can also decay to Higgs and longitudinal gauge bosons with a
coupling proportional to $\epsilon y_+$ in eq. (\ref{HH}). The competition between these two channel is a distinctive feature. Explicitly,
\be\label{wwlong}
\Gamma(\pi_0\to W_L W_L)=2\Gamma(\pi_0\to Z_L Z_L)=2\Gamma(\pi_0\to hh)=\frac{|y_{+}|^2 \epsilon^2}{32\pi}\frac{m_\rho^2}{m_\pi}
\ee
and similar rate for the charged components. Comparing these expression with the rate into photons we find
\begin{equation}
\label{wwratio}
\frac {\Gamma(\pi_0\to \gamma\gamma)}{\sum_V\Gamma(\pi_0 \to V_L V_L)}\sim   \frac{9\alpha^2}{16\pi^2} \frac {g_\rho^2}{\epsilon^2 y_+^2}\frac {m_{\pi}^4 }{m_\rho^4}
\sim\left(\frac {0.05}{\epsilon}\right)^2\left(\frac {0.05}{y_+}\right)^2 \left(\frac {2 m_\pi}{m_\rho}\right)^4
\end{equation}

Taking into account electro-weak symmetry breaking effects other interactions are possible. In particular $\pi$ inherits a coupling to SM fermions via the interactions of Eq.~\eqref{HH}, which gives a rate
\be
\label{wwferm}
\Gamma(\pi\to t\bar t) \sim 3 \frac{y_t^2}{16\pi} \epsilon^2 \frac{|y_+|^2}{g_\rho^2}\frac{v^2}{f^2}\frac{m_\rho^4}{m_K^4}m_\pi
\ee
This coupling is then important for the possible single production of the triplet which can be expressed as
\begin{equation}\label{eq:gg-pi0}
\Gamma(\pi^0\to gg)\approx \epsilon^2 \frac{|y_+|^2}{g_\rho^2}\frac{v^2}{f^2}\frac{m_\rho^4}{m_K^4} \Gamma(H\rightarrow gg)|_{m_H=m_\pi}.
\end{equation}
All in all, the dominant decay mode for the $\pi$ triplet will be determined by the relative size of the contribution of Eq.~\eqref{wwlong}, Eq.~\eqref{wwferm} and the anomaly contribution of Eq.~\eqref{anomalies}.
The anomalous decay to electro-weak gauge bosons dominates only for very small mixing, {\emph i.e.} when the Higgs is mostly elementary.

In general the electro-weak multiplets are split by the electro-magnetic interactions and for the case of the $\pi$ triplet the splitting turns out to be $\sim166\;$MeV~\cite{Cirelli:2005uq}. 
In addition the charged and neutral components of a real triplet can be also split by the effective operator
$
\frac{\epsilon^4 m_K^2}{f^4}\pi^a \pi^b H^\dagger \sigma^a H H^\dagger \sigma^b H
$
which is generated by the potential, but this contribution is smaller than 166 MeV.

\subsubsection{Tree-level regime: $y_+ y_- \lesssim 1$}\label{tree-level-regime}
When $y_+ y_- \lesssim 1$ we see from Eq.~\eqref{wwratio} and Eq.~\eqref{wwferm} that the decays to longitudinal modes and fermions dominate over the anomalous decays. Note that in this regime the branching fractions of $\pi_0$ into $ZZ$ do not depend on $y_{+}$ and $\epsilon$ but only on the details of the spectrum
\be\label{piZZ}
\mathrm{BR}(\pi_0 \to Z_LZ_L) \approx \frac{25\%}{1 + 6 y_t^2 v^2 m_\pi^2/m_K^4}.
\ee
On the contrary the production cross section is sensitive to $y_+^2\epsilon^2$ as per Eq.~\eqref{eq:gg-pi0}. In this case one can recast searches for resonances decaying to $ZZ$~\cite{CMSZZ} and $WW$~\cite{Aaboud:2017rel}. By fixing a mass ratio $m_\rho=3m_\pi=3m_K$ we obtain a limit on $\epsilon y_+$ of the order of 10$^{-1}$. This type of signature is correlated with the one that we will discuss in Sec.~\ref{sec:K-Pinv}, where a single produced $K$ decays to $h/Z \eta$ and current searches can give a bound  stronger than the one achieved for the triplet.

\subsubsection{Anomalous regime: $y_+ y_- \ll 1$}\label{3gamma}

\begin{figure}[t!]
\begin{center}
\includegraphics[width=0.44\textwidth]{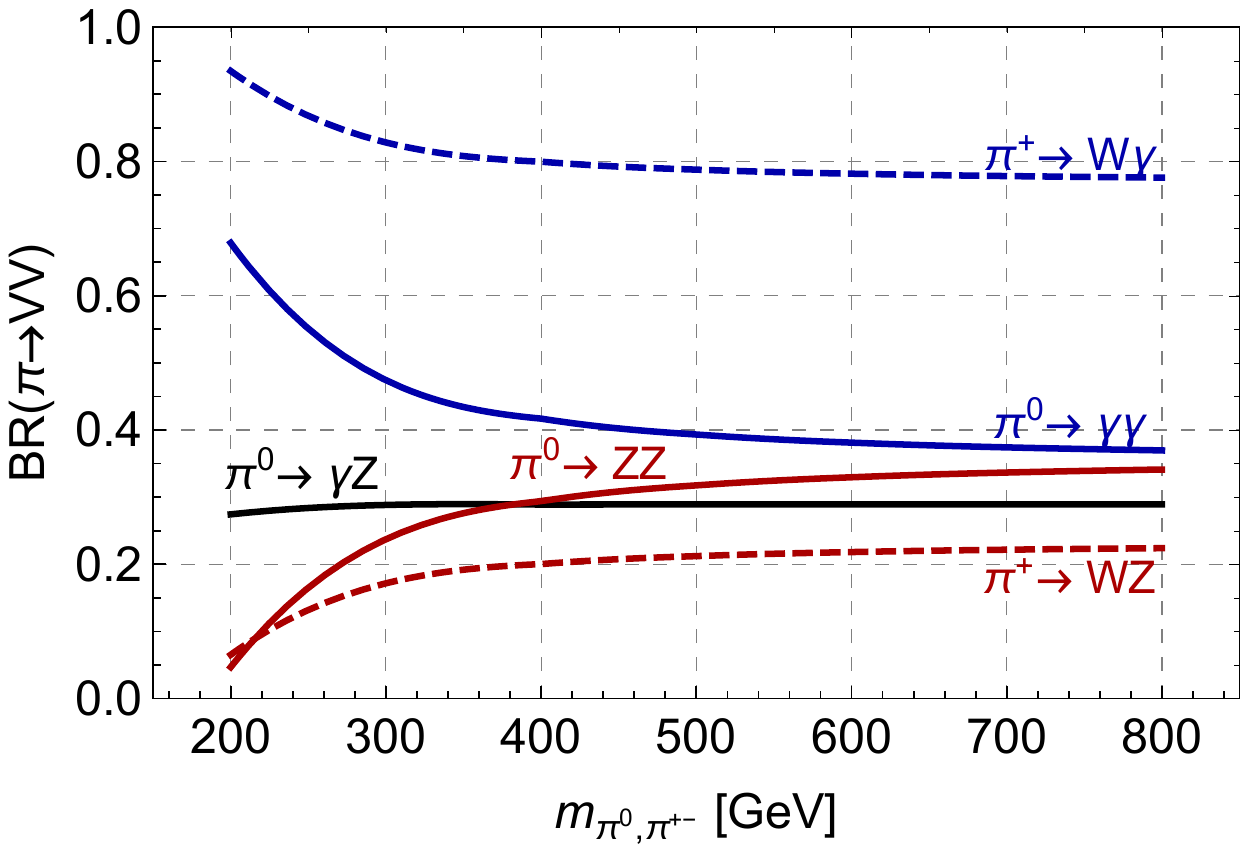}{}\hfill
\includegraphics[width=0.465\textwidth]{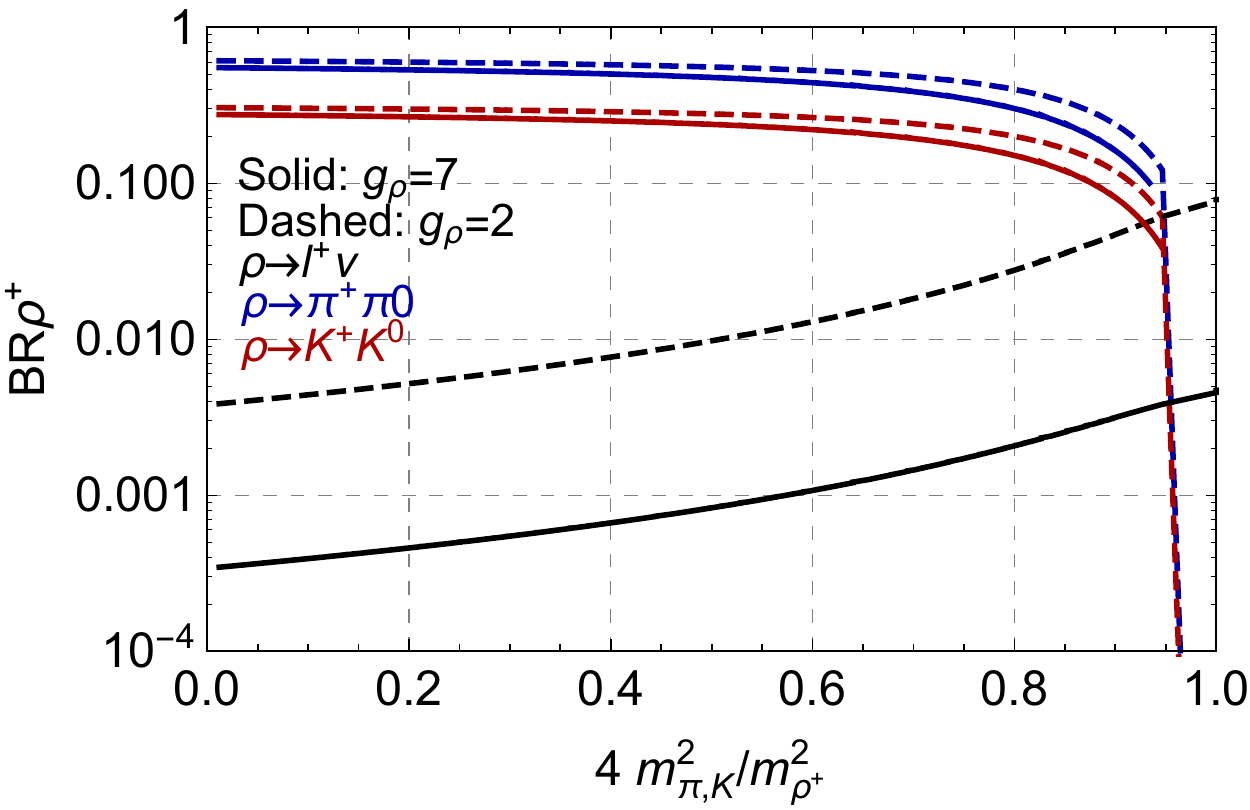}{}\hfill
\caption{\em Branching ratios of the neutral and charged $\pi$ into a pair of SM gauge bosons in the mostly elementary regime (left).
 Branching ratios of the charged $\rho$ into a pair of $\pi$, $K$ and SM leptons. $m_\pi=m_K$ is assumed in the plot (right).}
\label{fig:pion_rho_decay}
\end{center}
\end{figure}

When $y_+ y_- \ll 1$ the anomalous decays dominate over the other decay modes, and the branching ratios for $\pi^\pm$ and $\pi^0$ are shown in the left panel of Fig.~\ref{fig:pion_rho_decay}.
In this scenario the triplet can be produced with sizeable rates only through electo-weak interactions either in pair, via an s-channel exchange, or singly, via weak boson fusion. Since the former process is generically larger than the latter, in the following we will only discuss the pair production of the $\pi$ triplet states at the LHC.

\begin{figure}[t!]
\includegraphics[width=0.48\textwidth]{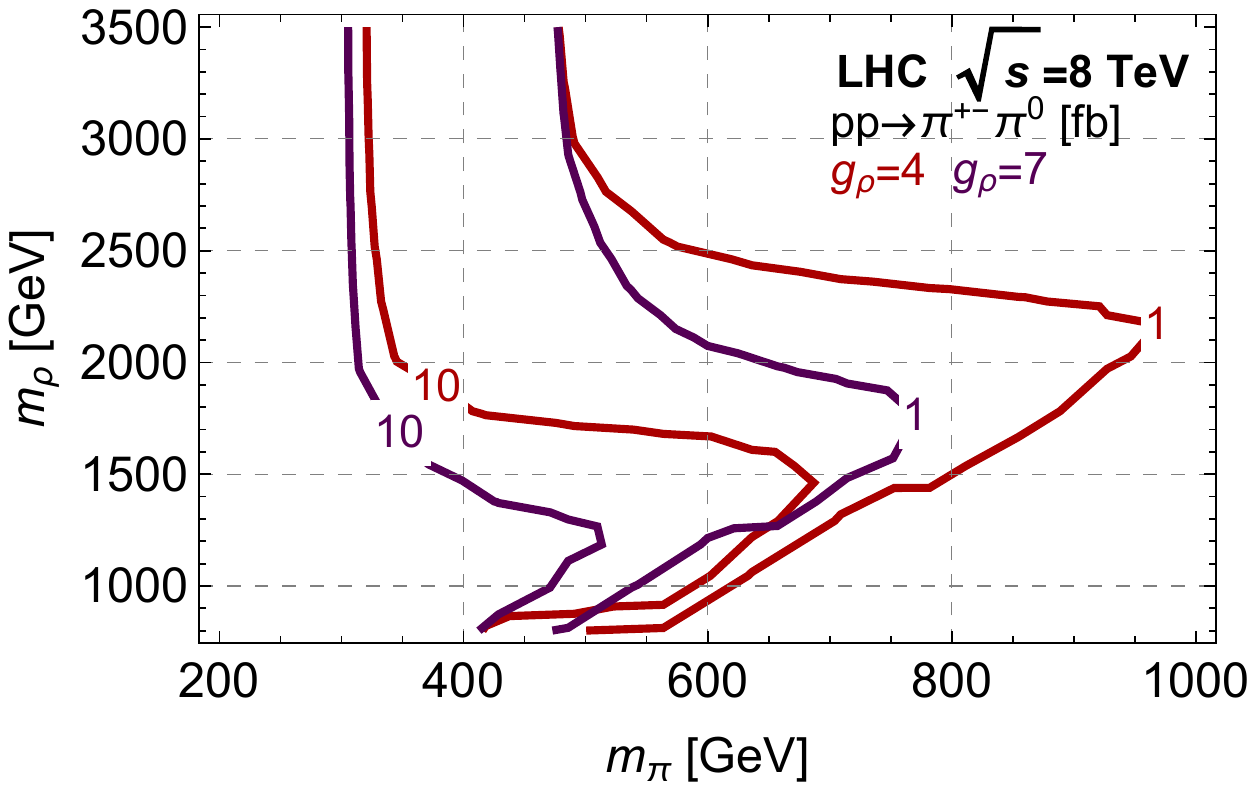}{}\hfill
\includegraphics[width=0.48\textwidth]{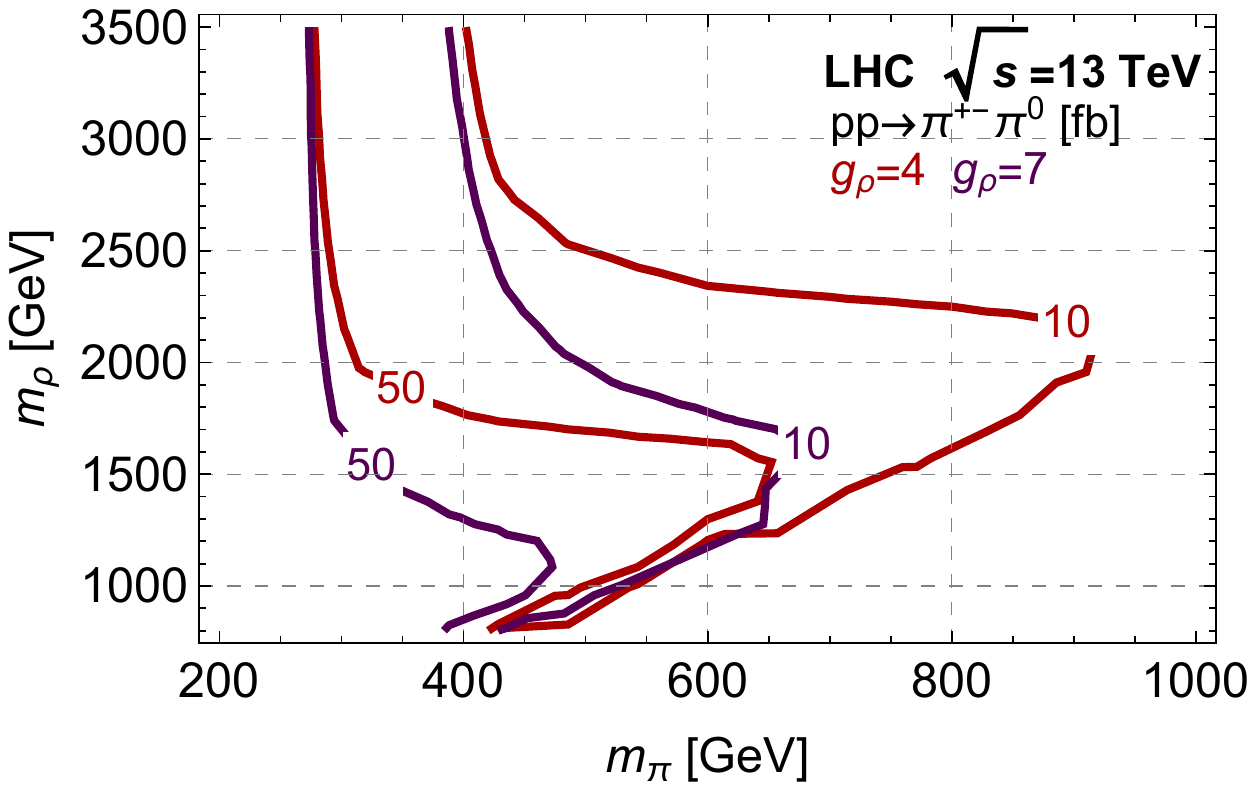}\hfill
\caption{\em Cross section for the production of a $\pi^{\pm}\pi^0$ pair in the $m_\rho-m_\pi$ plane for the LHC at $\sqrt{s}=$8~TeV (left) and 13~TeV (right).}
\label{fig:pion_xs}
\end{figure}

Pair production is mediated either through the exchange of a SM gauge boson or a $\rho$ resonance, where the latter process has a strong dependence on $g_\rho$, given the $g/g_\rho$ suppression of the resonance coupling to the SM quark current, see Eq.~\eqref{pheno}.
On the other side a larger $g_\rho$ increases the BRs of the $\rho$ into a  pair of triplets that, if kinematically allowed, tends to be the main decay channel for the heavy resonance, as illustrated in the right panel of Fig.~\ref{fig:pion_rho_decay}.
Thus, when $m_\rho > 2 m_\pi$, the triplet pair production cross section can be resonantly enhanced by the $\rho$ exchange. Taking into account both production modes the cross sections for the production of a $\pi^\pm\pi^0$ final state are illustrated in Fig.~\ref{fig:pion_xs}, where we have fixed $m_\pi=m_K$.

\begin{figure}[h!]
\begin{center}
\includegraphics[width=0.5\textwidth]{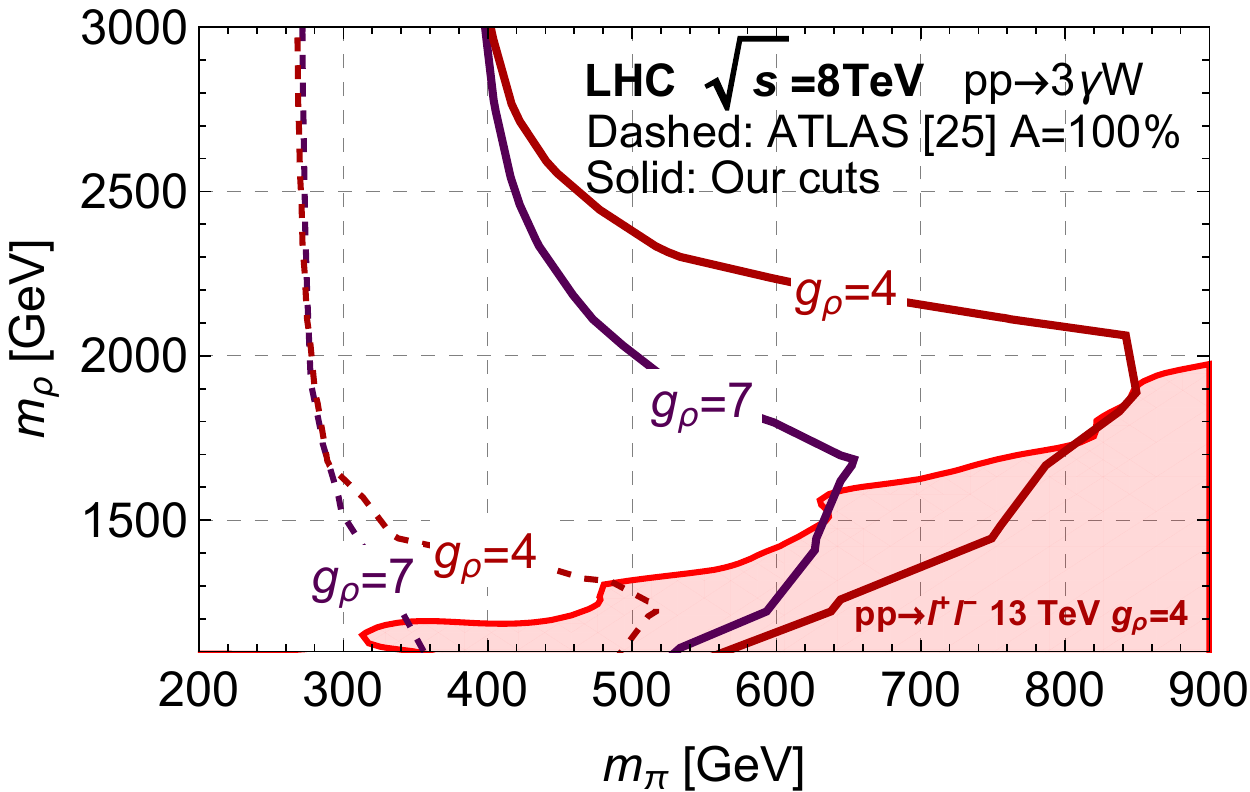}{}
\caption{\em 95\% CL exclusion in the $m_\rho-m_\pi$ plane for $g_\rho=4$ (red) and $g_\rho=7$ (purple). The solid lines represent the limit obtained with the proposed selection cuts while the dashed lines correspond the the limit obtained from the interpretation of the ATLAS analysis~\cite{Aad:2015bua} assuming an acceptance on the signal of 100\%. The shaded area shows  the region excluded by dilepton resonant searches~\cite{Aaboud:2017buh} with 36 fb$^{-1}$ of data. }
\label{fig:8TeV_res}
\end{center}
\end{figure}

Electro-weak pair production is extremely interesting for the triplet in the most elementary regime given the peculiar $\pi$ decay pattern that gives rise to a multiboson final state. 
Particularly interesting is the $3\gamma W$ signal which arises from the production of a $\pi^0\pi^\pm$ pair and which
is expected to be particularly clean at the LHC given the presence of three hard photons with a large transverse momentum set by the decaying $\pi$ masses. 
No dedicated analyses for a $3\gamma W$ signature exist.
However the ATLAS collaboration has performed an analysis at 8 TeV with 20.3 fb$^{-1}$ of integrated luminosity selecting events with 3$\gamma$ in the final state~\cite{Aad:2015bua}. Since they do not veto on the presence of extra activity, such as leptons and missing transverse energy, one can recast their results to set a bound on our scenario.
By assuming an ideal 100\% acceptance on the signal events, we obtain a bound $m_\pi \gtrsim  260\;$GeV when the $\rho$s are decoupled, limit that increases up to $m_\pi \gtrsim 500\;$ GeV when the $\rho$s can be resonantly produced. This is illustrated by the dashed lines in Fig.~\ref{fig:8TeV_res}. However the photons arising from the decay of the triplet are expected to carry a transverse momentum of the order of $m_\pi/2$, greater than the one required by the ATLAS analysis, which is roughly 20 GeV.
For this reason we expect that  the exclusion reach on this scenario could  be  improved.

\begin{figure}[t!]
\begin{center}
\includegraphics[width=0.45\textwidth]{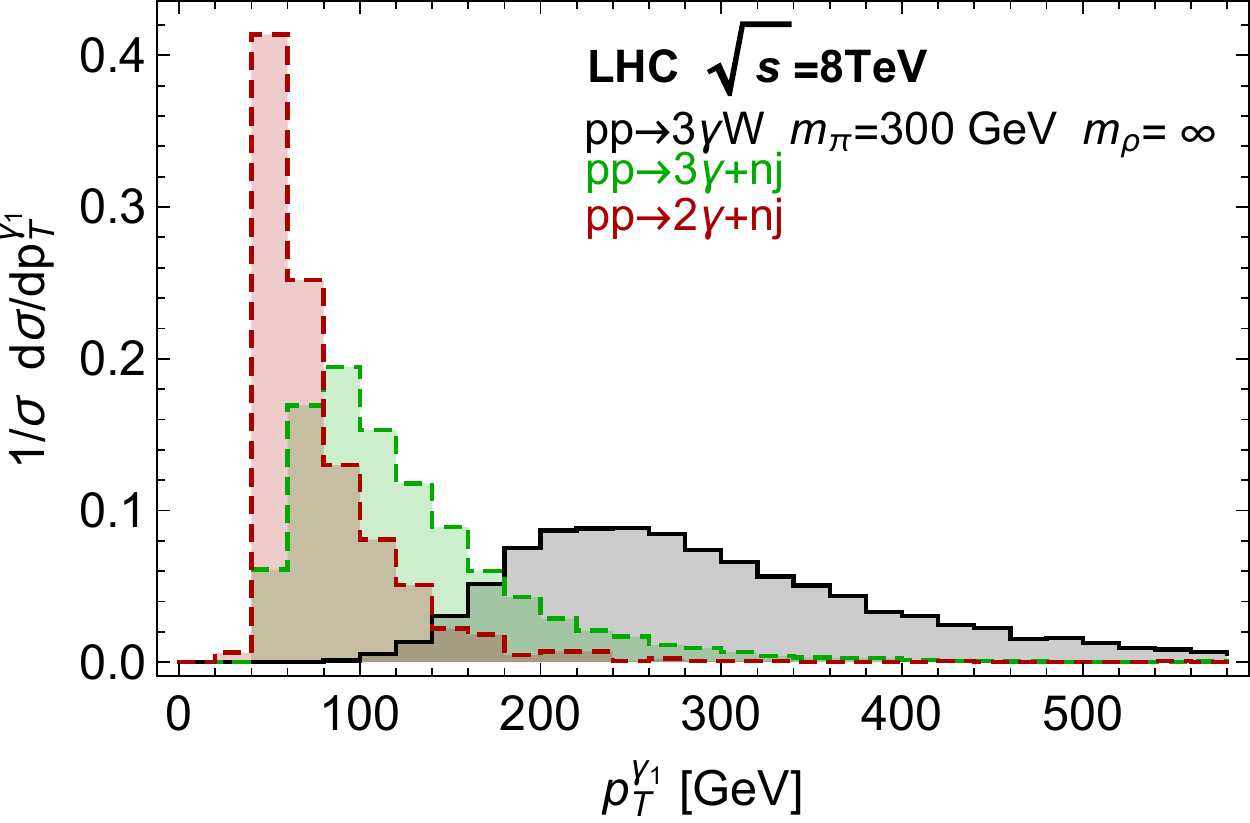}\quad
\includegraphics[width=0.47\textwidth]{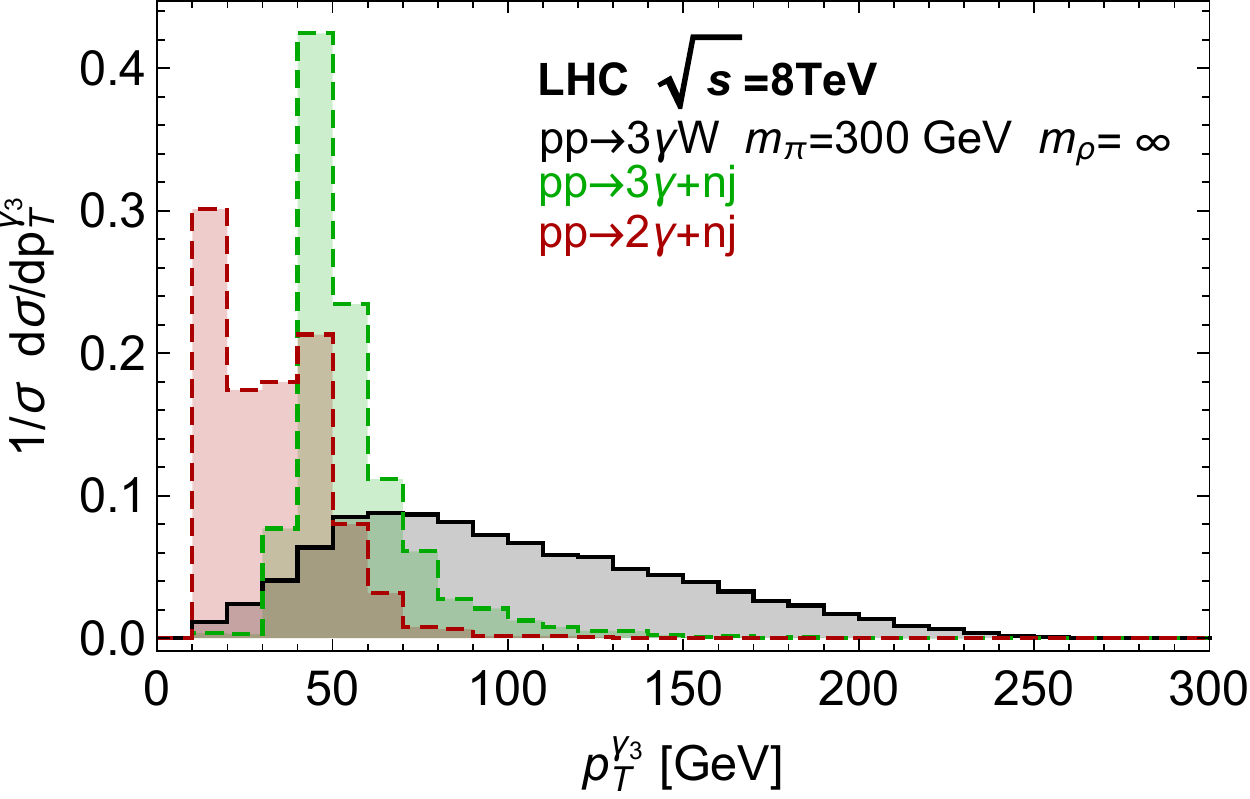}
\caption{\em Normalized distributions for the leading (left)  and third (right) photon transverse momentum for the signal (black), the $3\gamma$ background (red) and the $2\gamma j$ background (green). For the signal $m_\pi=$300~GeV and $m_\rho=\infty$ are assumed.}
\label{fig:8TeV_pt}
\end{center}
\end{figure}

In order to perform a dedicated analysis we have implemented\footnote{The model is publicly available at the {\tt FeynRules} web page \url{http://feynrules.irmp.ucl.ac.be/wiki/VLC_LN}.}  the lagrangian of Eq.~\eqref{pheno} in
the {\tt FeynRules} package~\cite{Alloul:2013bka} and
exported under the {\tt UFO} format~\cite{Degrande:2011ua}
 in order to make use of the {\tt
MadGraph5\_aMC@NLO} platform~\cite{Alwall:2014hca} for the simulation of hard scattering
LHC collisions.
Parton showering, hadronization  and decay of unstable SM particles have been performed via {\tt Pythia 8}~\cite{Sjostrand:2007gs} while we have simulated the response of the LHC detectors through the {\tt Delphes 3}~\cite{deFavereau:2013fsa} package. 
We simulate the signal and the main SM backgrounds for the $3\gamma$ final state, which are real $3\gamma + nj$ processes as well as $2\gamma + nj$ processes where additional photons arise from initial/final state radiation effects (ISR/FSR) and jet mis-identified as photons.  We refer to Appendix~\ref{appendixB} for the details on the validation of the background simulation. We then show in Fig.~\ref{fig:8TeV_pt} the transverse momentum distributions of the leading and third photon for the two main sources of background and the signal, the latter with $m_\pi=300\;$GeV and the $\rho$ decoupled.
From the figures is clear that the signal and the backgrounds peak at different values for the photon transverse momenta. We further notice that the fake background presents a softer spectra for the third photon $p_T$, which is mainly due to ISR/FSR photons and which can be then removed by asking the photon to have enough $p_T$. The peak at $p_T^{\gamma_3}\sim\;$50 GeV is instead due to a jet mis-identified as a photon and it peaks roughly at the same value of the real background as expected.

All together, we find that the reach\footnote{We have defined the statistical significance as $z=\frac{S}{\sqrt{S+B+ \alpha^2 B^2}}$ where $S$ and $B$ are the number of signal and background events respectively and $\alpha$ is the systematic error on the background determination, assumed to be 10\%. We have further multiplied our background estimates by a factor 2 to take into account for cross section determination uncertainties due to NLO effects.} on the $\pi$ mass is maximized for simple cuts on the transverse momenta of the three selected photons, namely $p_T^{\gamma_1,\gamma_2,\gamma_3}>250, 75, 75\;$GeV and we show our results in Fig.~\ref{fig:8TeV_res} for the choices of $g_\rho=4$ and 7. Here the solid lines represent the bound obtained with our analysis,  
to be compared with the dashed ones arising by recasting the ATLAS search~\cite{Aad:2015bua}.
Also shown in the plot is the region excluded by the 13 TeV ATLAS search for resonances in the dilepton final state with 36 fb$^{-1}$~\cite{Aaboud:2017buh} which is not visible for $g_\rho=7$ due to the reduced $\rho$ production cross section.
We see that already with existing 8 TeV data the reach on the $\pi$ mass in the regime where the $\rho$s are decoupled can be greatly increased, with bounds that can reach $\sim$ 600 (800) GeV for $g_\rho=7$ (4) when the $\pi\pi$ cross section is resonantly enhanced by the $\rho$ exchange. We also notice that the limits we obtain nicely complement the region already excluded by high mass dilepton searches, which are only effective in the region where $m_\rho<2 m_\pi$.

\begin{figure}[t!]
\begin{center}
\includegraphics[width=0.45\textwidth]{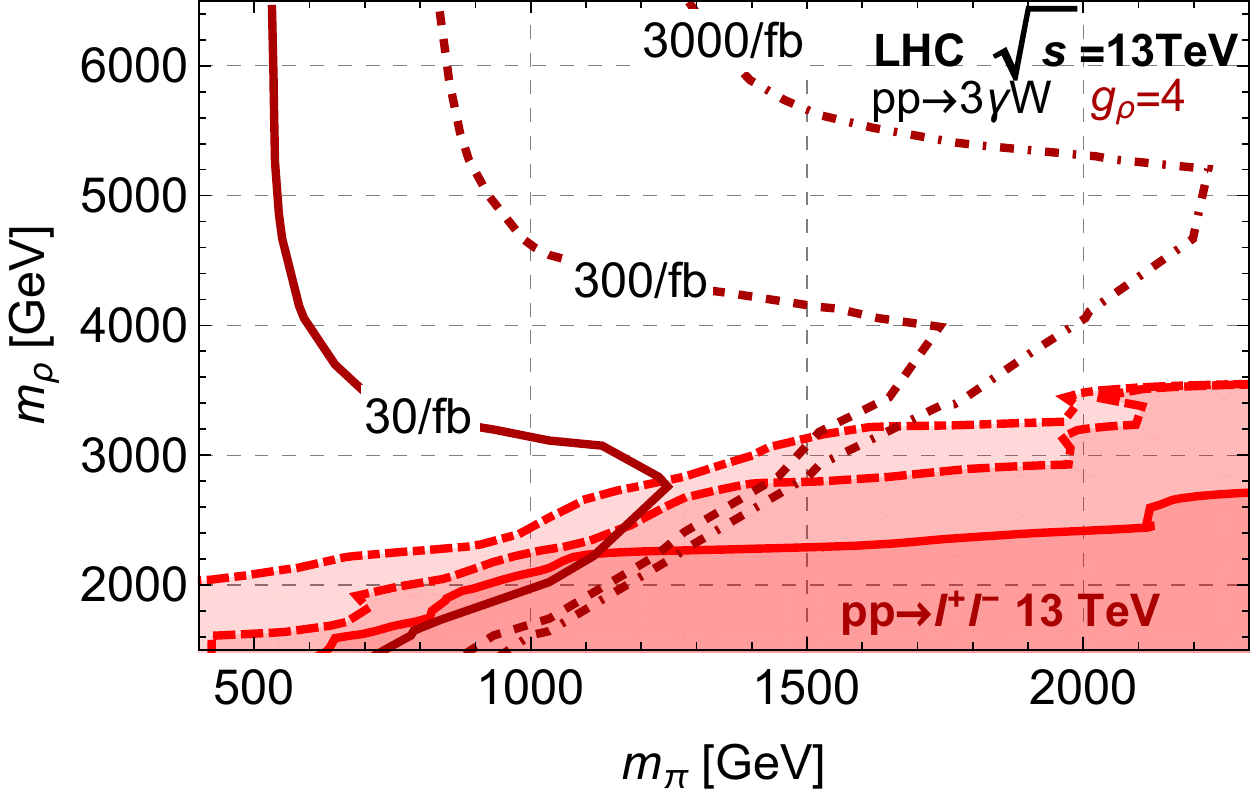}\quad
\includegraphics[width=0.45\textwidth]{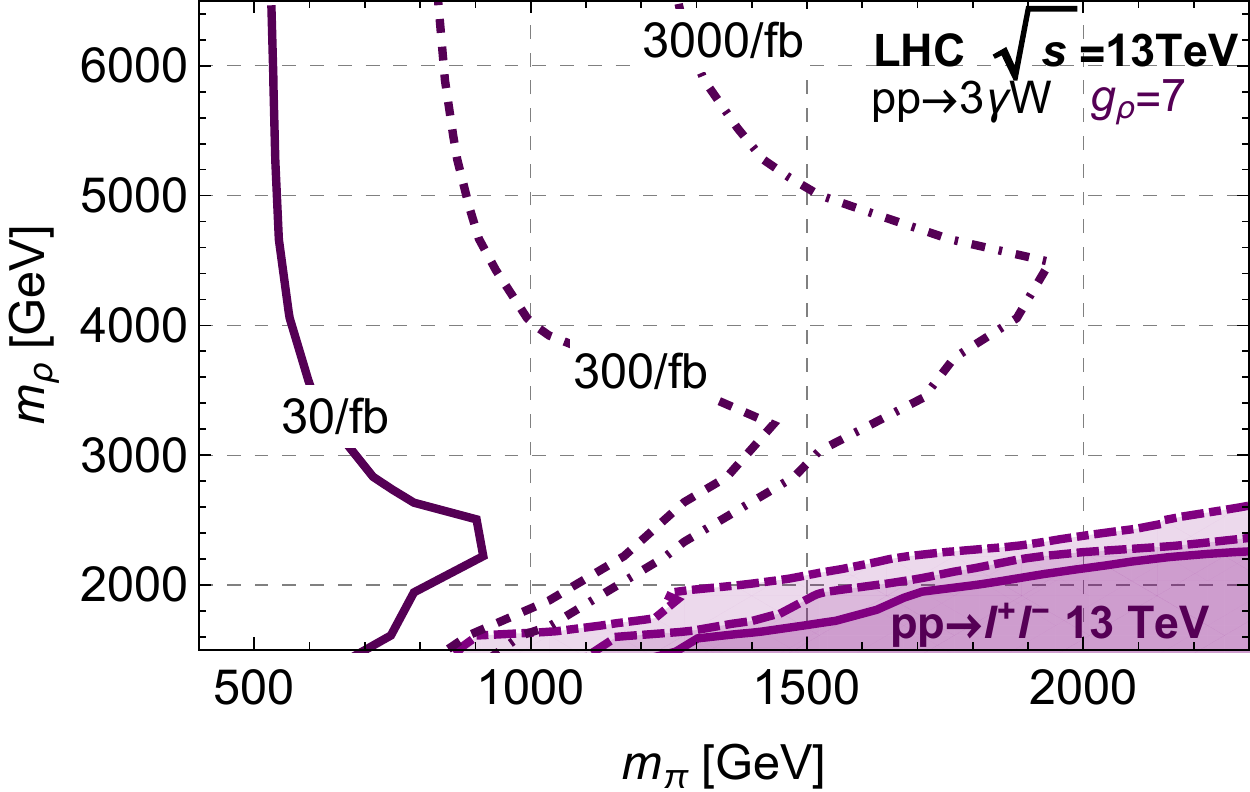}
\caption{\label{Fig:13}\em 95\% CL exclusion in the $m_\rho-m_\pi$ plane for $g_\rho=4$ (left) and $g_\rho=7$ (right) for 30, 300 and 3000 fb$^{-1}$ of integrated luminosity. The shaded areas correspond the the region excluded by dilepton resonant searches~\cite{Aaboud:2017buh}.}
\end{center}
\end{figure}

Given that the background is efficiently reduced by asking for photons with a large transverse momentum, we can expect that the current run of the LHC  at $\sqrt{s}=13\;$TeV can greatly improve on the results that can already be obtained with 8 TeV data.  We find that the exclusion reach in the $m_\rho-m_\pi$ plane is optimized for cuts on the leading, second and third photon transverse momenta of 300, 100 and 100 GeV for an integrated luminosity of 30 fb$^{-1}$ and 400, 300 and 300 GeV for integrated luminosities of 300 and 3000 fb$^{-1}$, Our results, together with the present and projected exclusion~\footnote{
The projected exclusion are computed by rescaling the upper bound on the dilepton cross section by $\sqrt{{\cal L}_0/\cal L}$, where ${\cal L}_0$ is the integrated luminosity of the ATLAS search~\cite{Aaboud:2017buh} and ${\cal L}$ is the target integrated luminosity.} from dilepton searches, are illustrated in Fig.~\ref{Fig:13} again for the case $g_\rho=4$ and 7.
All together wee see that the proposed analysis will be able to exclude, at the end of the high luminosity phase of the LHC, triplet mass up to 1.2 TeV in the non resonant regime, limits that can be pushed up to $\sim 1.8$ (2.2) TeV for $g_\rho=7$ (4) in the resonant case.


From the previous analysis we can also envisage  what are the main signatures for the detection of the spin-1 $SU(2)_L$ triplet at the LHC. When its decays into a pair of pions is kinematically closed, the most stringent bounds arise  from dilepton resonance searches. In the opposite regime, the most promising scenario is in the anomalous regime:  $y_+ y_- \ll 1$ , where one can exploit the clean anomalous decay of the $\pi$s to indirectly constraint the $\rho$ mass. On the other side, in the tree level regime ($y_+ y_- \lesssim 1$), one expects the production of 4 transverse weak gauge bosons with a rich, and complicated, final state.  Notice that when the decay to pNGBs is allowed, the $\rho$ tends to deviate from the narrow width approximation, which is very well motivated when such decays are kinematically closed. In particular we have that
$\Gamma_{\rho^+}/m_\rho= c\, g_\rho^2/(96\pi)$ where $c \gtrsim1$ and $c=1$ when only the triplet is considered.

\subsection{$K$ doublet}

The composite $K$ doublet would be stable due to species number conservation in the confining sector and in this case it would manifest itself either as charged tracks in the detector, for which current bounds are of the order of 400 GeV~\cite{CMS:2016ybj}, or as missing transverse energy. Its  decays are instead controlled by the elementary couplings $y_\pm$. Before electro-weak symmetry breaking we identify two main effects which depend on the relative size of $y_-$ and $y_+$. First, the mixing of $K$ with the SM Higgs doublet proportional to $y_-$ generates a coupling of $K$ with SM states with the usual pattern of a type-I two Higgs doublet model.
These couplings, decomposed in scalar, pseudo-scalar and charged components as $K=(K^+, \frac{K^0+i A}{\sqrt{2}})$, are related to the ones of a SM Higgs with the same mass as
\begin{align} 
&\frac{ g_{K^0ff}}{g^{SM}_{hff}}\bigg|_{\rm type-I} =\sin\delta +\frac{\cos\delta}{\tan\beta},\quad  &\frac{ g_{K^0VV}}{g^{SM}_{hVV}}\bigg|_{\rm type-I}=\sin\delta, \label{K-coupling} \\ 
&\frac{ g_{Auu}}{g^{SM}_{huu}}\bigg|_{\rm type-I} =\frac{1}{\tan\beta},\quad  &\frac{ g_{Add}}{g^{SM}_{hdd}}\bigg|_{\rm type-I}=-\frac{1}{\tan\beta}, \label{A-coupling}\\
& \frac{ g_{K^\pm qu}}{g^{SM}_{hqu}}\bigg|_{\rm type-I}=\frac{1}{\tan\beta}, & \frac{ g_{K^\pm qd}}{g^{SM}_{hqd}}\bigg|_{\rm type-I}=-\frac{1}{\tan\beta},\label{Kc-coupling}
\end{align}
and for the $K^0$ component, at leading order in $\epsilon$ and for $m_K\gg m_h$  they read
\be\label{K-fermion}
\frac{\delta g_{K^0ff}}{g_{hff}}\approx |\epsilon| ,\quad\quad \frac{\delta g_{K^0VV}}{g_{hVV}}\approx |\epsilon| \frac{m_h^2}{m_K^2}.
\ee
The couplings of Eq.~\eqref{K-coupling}-\eqref{Kc-coupling} allows $K$ to decay to SM fermions. 
For example, for the third generation quarks one finds\footnote{Here and in what follows we neglect final state masses that can significantly modify our estimates for $m_K< 500$ GeV.}
\begin{equation}\label{Kc-fermion}
\Gamma(K^+\to t \bar{b})= 2\Gamma(K^0\to t \bar{t})=2\Gamma(A\to t \bar{t})= \epsilon^2 \frac {3 y_t^2} {8\pi} m_K\,.
\end{equation}
Secondly, the presence of $y_{+}$ controls the decay of $K$ into $H$ and a singlet, with a decay rate given by
\begin{equation}\label{K-Heta}
\Gamma(K^\pm \to W^\pm_L + \eta)=2\Gamma(K^0 \to Z_L + \eta)=2\Gamma(A \to h + \eta)= \frac {|y_+|^2}{48 \pi}\frac {m_\rho^2}{m_K} ,
\end{equation} 
in the limit $m_\eta=0$. Note that despite the Yukawa couplings violate CP in general, due to the approximate degeneracy between the imaginary and real part of the neutral component of $K$, one can unambiguously identify $K^0$ and $A$. The amplitude for $K\to H + \pi$ is parametrically similar and differs only by a calculable group theory factor, however we do not expect a sizeable decay width since $m_K\sim m_\pi$ while the singlet can be much lighter.
Therefore,  neglecting  electro-weak symmetry breaking effects, the relevant decay mode of $K$ is determined by the competition between Eq.~\eqref{K-fermion} and Eq.~\eqref{K-Heta}.  One finds 
\be\label{ratio-K}
\frac{\Gamma( K^0 \to t \bar{t})}{\Gamma(K^0 \to H \eta)}\sim 18 \epsilon^2 \frac{y_t^2}{|y_+|^2}\frac{m_K^2}{m_\rho^2}\sim 36 \frac{y_-^2}{|y_+|^2}\frac{y_t^2}{g_\rho^2}\frac{m_\rho^2}{m_K^2},
\ee
where in the second equality we used the explicit expression for $\epsilon$, see Eq.~\eqref{eq:mixing}. This equation shows that when $\epsilon/y_+\ll 1$, or analogously $y_-/y_+\ll 1$,  the decay into $H\eta$ dominates.

After electro-weak symmetry breaking other effects take place. Allowing for $\langle K^0 \rangle \approx \epsilon v$ in the non-linear derivative interactions of the NGBs, one obtains the term $\epsilon v K (\partial \eta)^2/f^2$, which is of the same order of the $\epsilon m_K^2 K^\dag H \eta^2/f^2$ term arising from the expansion of the potential, that give a decay rate for $K$ into a pair of singlets
\be
\Gamma( K^0 \to \eta \eta)\sim \frac{\epsilon^2}{1152\pi}\frac{m_K^3}{f^2} \frac{v^2}{f^2}
\ee
which is a sizable effect only in the mostly composite regime \cite{Gripaios:2009pe}. In general, when CP is broken, all the neutral scalars will mix with each other and this induces corrections to the couplings used to derived the above quantities, introducing an uncertainty of $O(y_{+}^2g_\rho^2 v^2 f^2/m_K^4)$. We also note that a mass splitting between the $K$ neutral and charged components is induced by the quartic interaction $\epsilon^2 m_K^2/f^2  K^\dag \sigma^a K H^\dag \sigma^a H$ leading to $\Delta m_K/m_K \sim \epsilon^2 v^2/f^2$
leading to sizable splittings for $\epsilon> 0.1$.


Because of the coupling to the top quark $K^0$ and $A$ can be singly produced via gluon fusion similarly to the SM Higgs with rates 
\begin{equation}
\Gamma(K^0\to gg)\approx\Gamma(A\to gg)\approx \epsilon^2 \Gamma(H\rightarrow gg)|_{m_H=m_K} 
\end{equation}
More precisely, the rates of the scalar and pseudo-scalar components are equal in the high-energy limit $m_K\gg 2m_t$, with the pseudo-scalar one being larger by factor of $(3/2)^2$ in the opposite  regime.  While, barring the rescaling due to the different loop structures, the production is rather insensitive to the CP properties of the neutral $Ks$, the decay depends upon it. For example in the case $y_+ \gg y_-$ the neutral $K^0$ decays into $h\eta$ while $A$ decays into $Z\eta$. On the other hand   the decay into pairs of SM fermions is still rather insensitive to the CP structure.  Another mechanism of production for the $K$ is double production via electro-weak interactions that can give rise to interesting signatures. Nevertheless we expect them to be less clean  and we do not study them further, focusing on the phenomenology of singly produced $K$. 

\subsubsection{Almost $P$ invariant regime: $y_- \ll y_+$}
\label{sec:K-Pinv}

The composite $K^0$ entirely decays into a SM Higgs boson and the $\eta$ singlet.
This is an interesting decay mode since, through the anomalous decay of the singlet $\eta$, it can give rise to a $2b2\gamma$ final state, a signal targeted by double Higgs production searches~\cite{Aad:2014yja,Khachatryan:2016sey,TheATLAScollaboration:2016ibb,CMS:2017ihs} that can be reinterpreted when $m_K>m_\eta+m_h$. 
Note that the term of Eq.~\eqref{HH} gives rise, if kinematically allowed, to a $\eta \to V_L V_L$ rate analogous to the one of the triplet, Eq.~\eqref{wwlong}, thus reducing the $\eta \to \gamma\gamma$ rate. We thus focus on a light $\eta$. Moreover, by comparing Eq.~\eqref{ratio-eta} and Eq.~\eqref{ratio-K}, one sees that when the anomalous decay $\eta\to \gamma\gamma$ dominates, the same can happen for the process $K\to H \eta$.

\begin{figure}[t!]
\begin{center}
\includegraphics[width=0.45\textwidth]{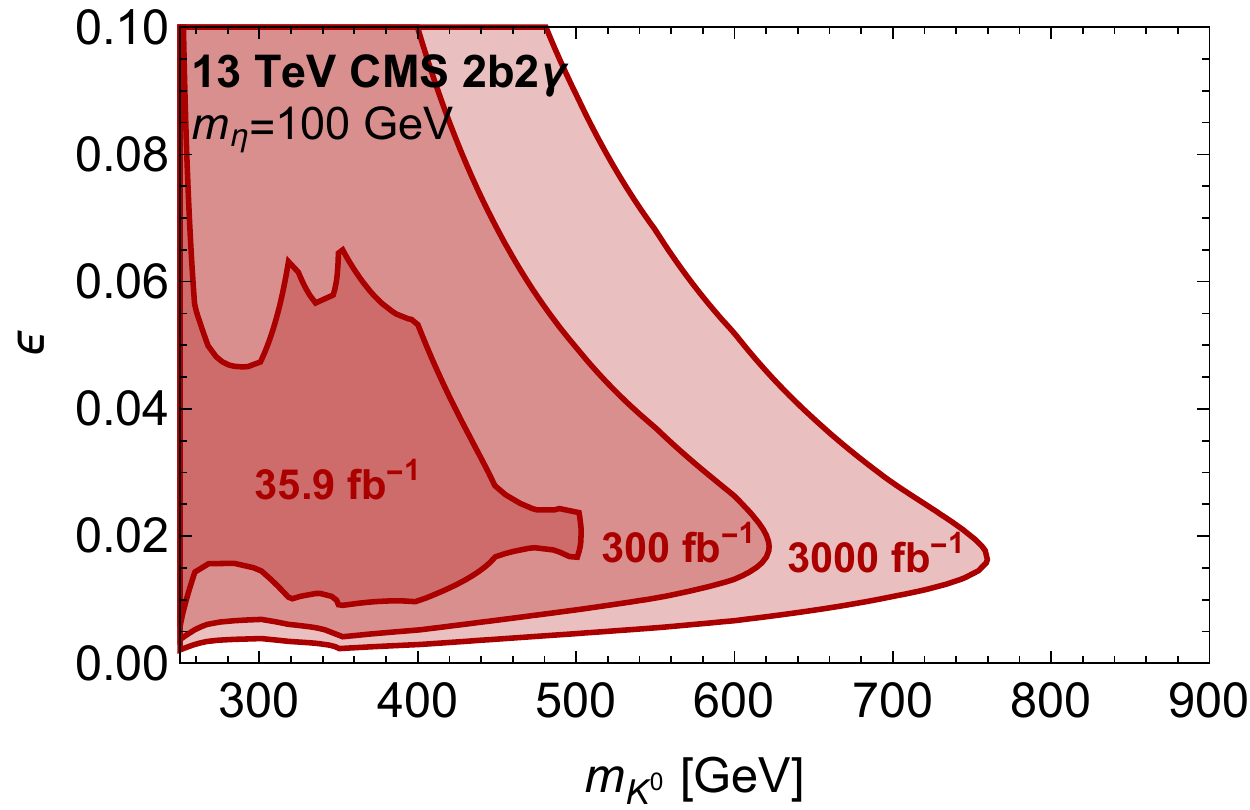}
\caption{\label{fig:kappa} \em 95\% CL exclusion limit in the $m_{K^0}$-$\epsilon$ plane for $m_\eta=$100~GeV. The plot is shown requiring  $\Gamma(K\to h \eta)=2\Gamma(K\to t\bar t)$ as per eq.~\eqref{ratio-K}, which then fixes the branching ratio for $\eta\to \gamma\gamma$ in all the regions of the plot (for $m_\rho=2\ \mathrm{TeV}$ and $g_\rho=7$). The limits are obtained by reinterpreting the CMS analysis~\cite{CMS:2017ihs}, see the main text for the details.}
\end{center}
\end{figure}

Interestingly, while in the SM the branching ratio of the Higgs boson in a $2\gamma$ final state is of the order of $10^{-3}$, the singlet rate into a diphoton final state is of order unity for $m_\eta<160\;$GeV, if only the anomalous decay is present, and decreases down to $\sim\;$0.25 for $m_\eta=250\;$GeV. This difference translates into a large rescaling of the experimental upper limit on the di-Higgs cross section which in turn allows to set strong bounds on $\epsilon$. These limits, obtained by using the CMS analysis~\cite{CMS:2017ihs}, are shown in Fig.~\ref{fig:kappa} for the presented dataset of and also extrapolated~\footnote{We extrapolate to higher integrated luminosities by using the expected limit set by the analysis.} for higher integrated luminosities for $m_\eta=100$ GeV. We see that limits on $\epsilon$ of order of $10^{-2}$ can already be set with the present dataset with these limits becoming slightly stronger (weaker) for heavier (lighter) $m_\eta$ mass. These limits are expected to improve by a factor 2 at the end of the high-luminosity phase of the LHC.

A comment here is in order. While reinterpreting the experimental limits we are assuming that the signal selection efficiency on the $h\eta\to 2b2\gamma$ final state is the same as for the $hh\to 2b2\gamma$ final state,  in our scenario the $2\gamma$ invariant mass will be peaked at the $\eta$ mass, generally different from $m_h$.
However we expect that by selecting a window for the invariant mass cut on the diphoton system of the same size as in the experimental analyses, but peaked now at the $\eta$ mass, the efficiency on the signal will approximatively be the same. 
Moreover, for $m_\eta>m_h$ this cut will reduce the background more than in the $m_\eta=m_h$ case. This is because the non-resonant backgrounds are a decreasing function of $m_{2\gamma}$ and the resonant ones (arising, {\emph e.g.},  from process of single Higgs production in association with vector boson or $tt$) will be almost completely removed. In this case our limits turn out to be conservative. 
On the other hand, in the regime with $m_\eta<m_h$, the resonant backgrounds will again be efficiently suppressed while the non resonant one will slightly increase.
We do not perform a dedicated background simulation for this final state. We however expect that the real limits will not be very different from the ones presented in Fig.~\ref{fig:kappa}.

\subsubsection{Tree-level regime: $y_- \gg y_+$ scenario}

When $y_- \gg y_+$ the $K$ main decay channels are into SM final states, with the rates inherited by the ones of the SM Higgs boson rescaled by the relevant couplings, see Eq.~\eqref{K-coupling} and Eq.~\eqref{A-coupling}. Note that $A$ doesn't couple to $WW$ and $ZZ$.

\begin{figure}[t!]
\begin{center}
\includegraphics[width=0.45\textwidth]{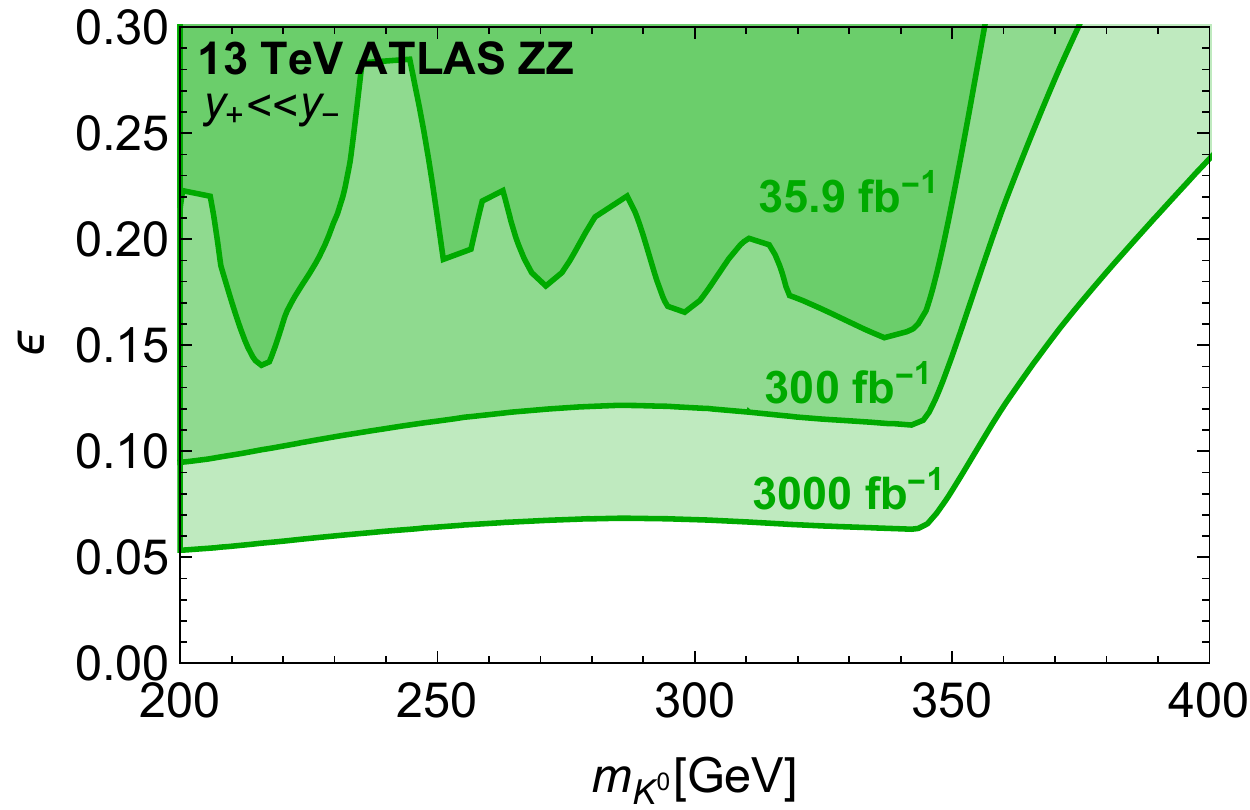}\quad
\includegraphics[width=0.45\textwidth]{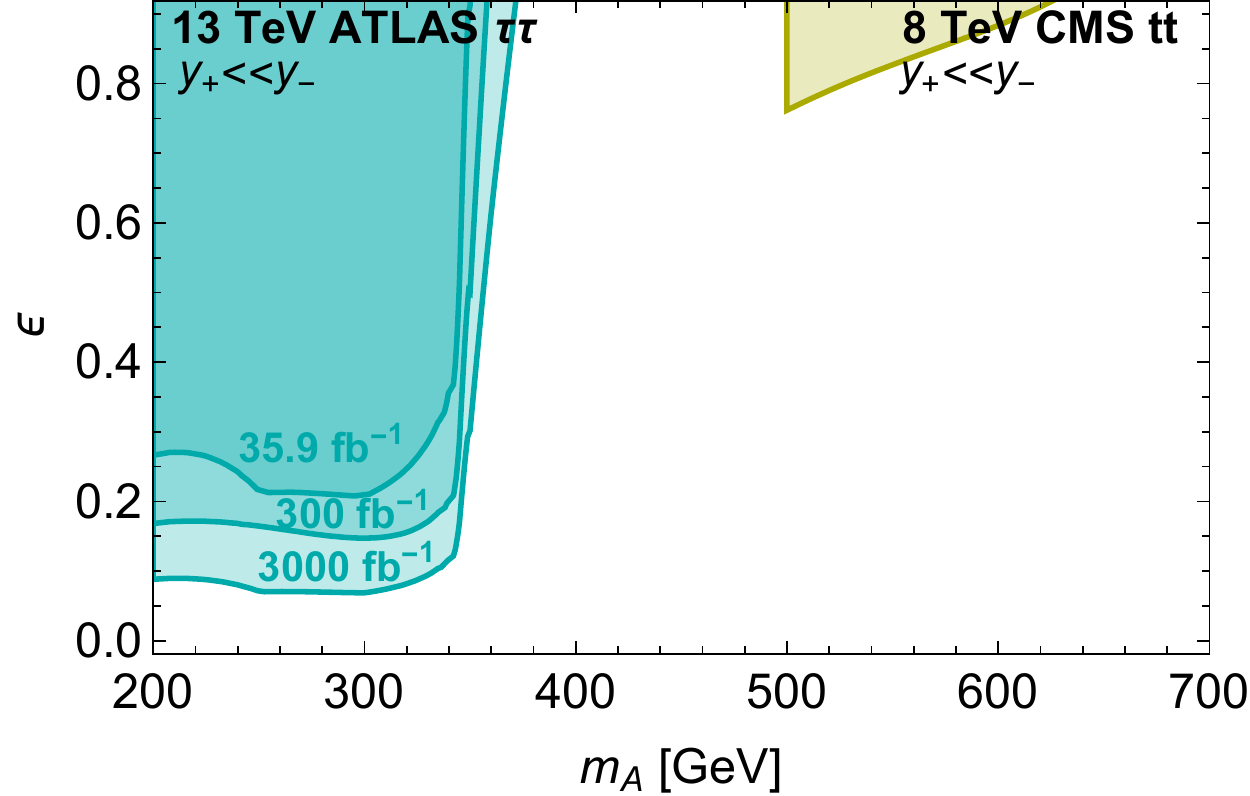}
\caption{\label{fig:kappa2}\em 95\% CL exclusion limit in the $m_{K^0}$-$\epsilon$ (left) and $m_A$-$\epsilon$ (right) plane from current LHC searches, see the main text for the details.}
\end{center}
\end{figure}

Above the $t\bar t$ threshold $K^0$ and $A$  will almost entirely decay into a pair of top quark. However, $t\bar t$ resonant searches are not yet sensitive. Below the $t\bar t$ threshold, $K^0$ decays into a pair of weak bosons, $WW$ and $ZZ$, dominate and one can recast LHC searches for $ZZ$ resonances carried out by ATLAS~\cite{Aaboud:2017rel} and CMS~\cite{CMSZZ}. We illustrate in Fig.~\ref{fig:kappa2} (left) the limits arising from the ATLAS analysis in the $4\ell$ and $2\ell 2\nu$ final state~\cite{Aaboud:2017rel} performed with an integrated luminosity of 36.1 fb$^{-1}$. As in the regime $y_- \ll y_+$ limits of order of $10^{-2}$ on $\epsilon$ can be enforced, which are expected to become roughly a factor 4 stronger with 3000 fb$^{-1}$ of integrated luminosity collected. We also note that below the diboson threshold $\tau\tau$ and $b\bar b$ decay modes become the dominant one but the rates are not yet within the LHC sensitivity.

The situation is different for the pseudo-scalar component $A$. Since it doesn't couple a tree level to $WW$ and $ZZ$, it's branching ratios are mostly into $tt$, $bb$ $\tau\tau$ and $gg$. In this case the strongest limits are given by $2\tau$~\cite{Aaboud:2017sjh} searches for $m_A<350$ GeV, while above the top pair threshold 8 TeV $t\bar t$ resonance searches~\cite{Khachatryan:2015sma} are only able to set a weak limit on $\epsilon$. We also notice that no limits can be set for $m_A$ between 350 GeV and 500 GeV.

\subsection{$\phi$ quintuplet}

Models with fermions forming a triplet $V$ of $SU(2)_L$ have new features. Among the  NGBs there are isospin triplet and quintuplet representations.
While the isospin triplet is stable because of $G$-parity~\cite{Bai:2010qg}, the quintuplet is unstable and decays though anomalies to electro-weak gauge bosons, see again Eq.~\eqref{anomalies}.
As for the triplets the different components of the quintuplet are split by electro-weak gauge loops so that
the neutral component is the lightest\footnote{The leading $SU(2)_L$ breaking interaction with the Higgs appear at dimension 6,
$
O=\phi_{ac} \phi_{cb} H^\dagger \sigma^a H H^\dagger \sigma^b H
$
and are  subleading.}.

The Drell-Yan production cross section of the quintuplet components is related to the one triplets as
\begin{eqnarray}
\sigma(pp\to \phi^{++}\phi^{--})=4\times\sigma(pp\to \phi^{+}\phi^{-})= 4\times \sigma (pp\to \pi^+ \pi^-)\,,\nonumber \\
\sigma(pp\to \phi^{\pm\pm}\phi^{\mp})=\frac{2}{3}\times\sigma(pp\to \phi^{\pm}\phi^{0})= 2\times \sigma (pp\to \pi^\pm \pi^0)\,.
\end{eqnarray}
The states $\phi^{\pm}\phi^0$ are produced with a rate a factor three larger than  $\pi^\pm \pi^0$ and decay with similar branching fraction to SM gauge bosons.
Differently from triplets the decay to $W_L, Z_L, H$ is suppressed so that the decay through anomalies always dominates.
In this case bounds similar, and stronger, to the ones of Fig.~\ref{fig:8TeV_res} and Fig.~\ref{Fig:13} therefore apply.

Also interesting is the production of doubly charge states, which gives rise to new signatures. Since $\phi^{++}$ can only decay into a pair of same sign $W$s one expect to obtain striking same-sign dilepton final states.
In this case one can recast experimental analyses targeting the production of doubly charged Higgs. 
In particular the 13 TeV ATLAS analysis~\cite{Aaboud:2017qph} searches for doubly charged Higgs bosons directly decaying into a pair of same-sign leptons.  By just taking into account the $W$ boson branching ratios in the process
\be
p p \to \phi^{++}\phi^{--} \to 2W^+ 2W^-
\ee
one obtains a bound of $m_{\phi^{++}}\gtrsim 400\;$ GeV by assuming a 100\% acceptance on the final state, bound that degrades down to $m_{\phi^{++}}\gtrsim 250\;$ GeV for an acceptance of $20$\%. These bounds are thus weaker than the one that can be obtained from $\phi^+\phi^0$ production.

\section{Conclusions}

The possibility of a strongly coupled sector lying at the TeV scale is  allowed by indirect constraints when it does not play a major role in the breaking of the electro-weak symmetry.
As shown in Fig.~\ref{fig:indirect} a degree of compositeness of $\sim 10\%$ allows us to easily satisfy precision constraints.

\begin{figure}[t!]
\begin{center}
\includegraphics[width=0.52\textwidth]{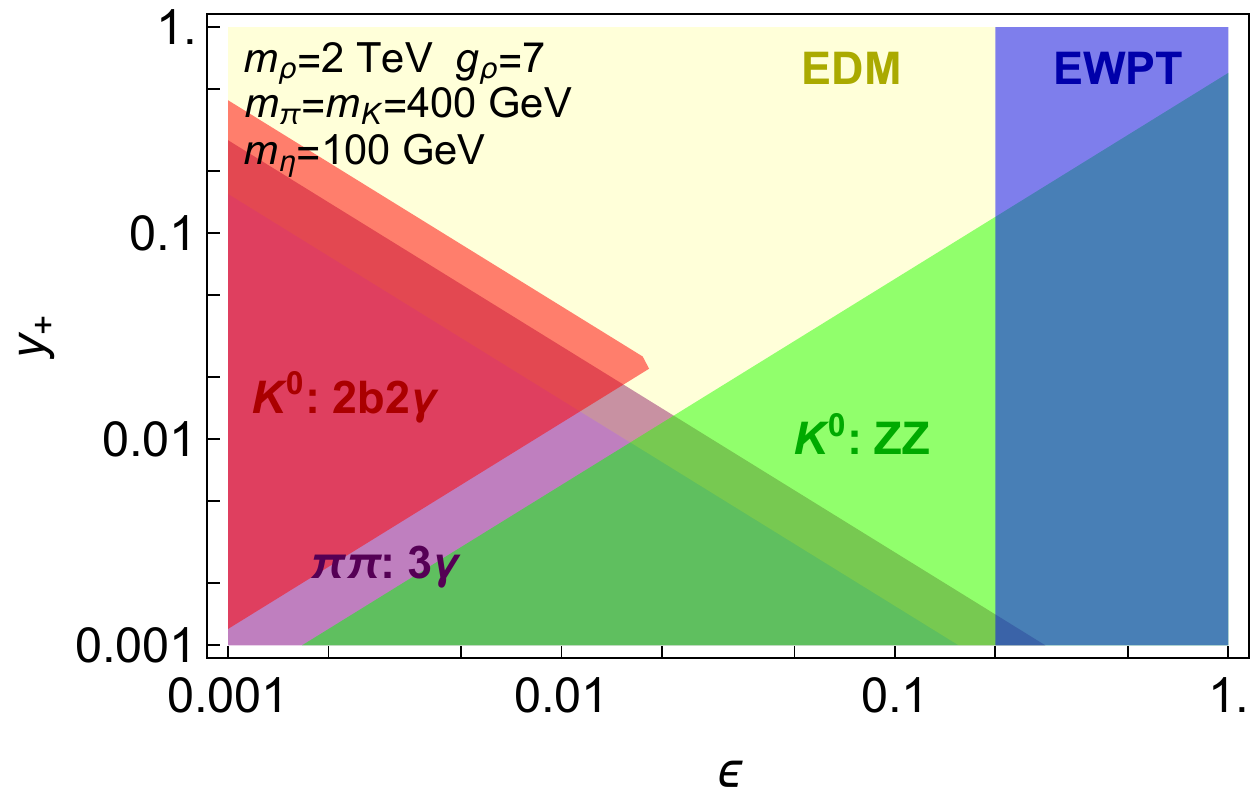}
\caption{\label{fig:summary} \em Relevant searches in different regions of parameter space. 
The EDM and EWPT constrain the yellow and blue regions, while the dominant collider signatures are depicted in red and green for the doublet and purple for the triplet.}
\end{center}
\end{figure}

On the other hand, new states with a $10\%$ mixture with the Higgs have a chance to be produced at colliders with sizeable rates. 
The phenomenology at the LHC is  dictated by the global symmetries of the strong sector and the quantum numbers of the vector-like fermions, that select the number of NGBs, their SM representation and, by a comparable amount, also by the Yukawa couplings $y_{-}$ (related to the mixing $\epsilon$) and $y_{+}$. We have found a rather rich structure depending on the size of these couplings, that we summarize here in three main categories, as well as in Tab.~\ref{tab:summary}.

\begin{table}[h]
\begin{center}
\begin{tabular}{c|c|c|c|c}
\multicolumn{5}{c}{Anomalous scenario}\\ \hline\hline
 NGB & Production & Decay & Model parameters & LHC \\ \hline
 $\pi$ & EW pair prod. & multi-$V_T$ & $c_{VV}N/f_\pi$ & {\huge \textcolor{DarkGreen}{\checkmark}}\\
 $K$ & EW pair prod. & disappearing tracks/HSCP/$E_T^{\rm miss}$ & -- & \textcolor{DarkGreen}{\checkmark}
\end{tabular}
\vskip 20pt
\begin{tabular}{c|c|c|c|c}
\multicolumn{5}{c}{Tree-level scenario  $y_{-}\gg y_{+}$}\\ \hline\hline
 NGB & Production & Decay & Model parameters & LHC \\ \hline
 $\pi$ & $gg$-fusion & $V_LV_L$ & $\epsilon y_{+}$  & \textcolor{DarkGreen}{\checkmark}\\
 $K$ &  $gg$-fusion  & $V_L V_L$ & $\epsilon$ & {\huge \textcolor{DarkGreen}{\checkmark}}\\
 $\eta$ & $gg$-fusion & $V_T V_T,tt,bb$ & $\epsilon y_{+}$  &
\end{tabular}
\vskip 20pt
\begin{tabular}{c|c|c|c|c}
\multicolumn{5}{c}{$P-$invariant scenario  $y_{+}\gg y_{-}$}\\ \hline\hline
 NGB & Production & Decay & Model parameters & LHC \\ \hline
 $\pi$ & $gg$-fusion & $V_L V_L$ & $\epsilon y_{+}$  & \textcolor{DarkGreen}{\checkmark}\\
 $K$ &  $gg$-fusion  & $H\eta$ & $\epsilon$ & {\huge \textcolor{DarkGreen}{\checkmark}}\\
  $\eta$ & $gg$-fusion / $K$ decay & $V_T V_T,tt,bb$ & $\epsilon y_{+}$  & \textcolor{DarkGreen}{\checkmark}
\end{tabular}
\caption{\em \label{tab:summary} Summary of the different regimes for production and decay of singlet, doublets and triplets. Here $V_L=(h,W_L,Z_L)$ and $V_T=(\gamma,W_T,Z_T)$. The presence of the checkmark shows that there are potential constraining searches from the  LHC, and the size of the checkmark within a given scenario is  an indication of the most promising channel to constrain the model. For the quantitative discussion see Sec.~\ref{collider}.}
\end{center}
\end{table}

\begin{itemize}
\item \textit{Anomalous scenario}.
When the Yukawa couplings are zero (or, say, totally negligible) the phenomenology is rather universal since the link with the SM is via gauge and anomalous interactions. Real NGBs decay through anomalies, with branching fractions that can be computed in terms of the quantum numbers of the vector-like fermions. Multi vector boson final states are rather promising, for example in the case of the EW production of triplets. In Sec.~\ref{3gamma} we recast existing ATLAS $3\gamma$ searches from Run I, and we study a dedicated search based on the idea that  a harder photon $p_T$ would help to reduce the background from fake jets and allow to explore kinematic configurations not analyzed in the original search, possibly doubling the mass reach. Complex NGBs on the other hand, would manifest themselves as charged tracks in the detector, for which current bounds are of the order of 400 GeV~\cite{CMS:2016ybj} or simply give rise to events with missing energy.
\item \textit{Mixed scenario}. For mixing $\epsilon\lesssim 0.1$, the phenomenological pattern change drastically, since tree-level processes tend to overcome the loop-sized rate of the anomalous terms. When this happens, then the major discriminant is the relative size between $y_{-}$ and $y_{+}$ couplings. As emphasized in Sec.~\ref{sec:symmetries} these control different (accidental) symmetries  and we can identify two regimes:
\begin{enumerate}
\item Case $y_{-}\gg y_{+}$. SM  final states generated at tree-level will be dominant and the quantitative predictions depend upon the quantum numbers of the NGBs. Indeed in this case  $K$, whenever allowed by phase space, will decay to third generation fermions (mainly $t\bar t$), and to (longitudinal) dibosons below the $t\bar{t}$ threshold and would be conspicuously produced via gluon fusion. The triplet instead decays mainly to (longitudinal) diboson and subdominantly to $t\bar t$, although since its coupling to fermions arise only after EWSB the production rate is smaller than the one for $K$.
\item Case $y_{+}\gg y_{-}$. The most interesting effects are related to the doublet $K$ which decays to final states with Higgs and an NGB, most likely a singlet, giving rise to the interesting final states $b\bar b \gamma\gamma$. By recasting existing searches for di-Higgs we were able to get strong limits on this scenario. 
Notice that this is also one of the few cases where the singlet $\eta$ can have a sizeable production rate.
\end{enumerate}
\end{itemize}

Despite the simplicity of the partially composite Higgs, fixed by the accidental symmetries of the fundamental gauge-Yukawa theory in the UV, the collider phenomenology explored in this work shows interesting patterns at the LHC. Together with the weak constraints arising from precision physics, this strengthen the motivation to explore at colliders this type of new physics scenario, which, despite being ``unnatural'', could plausibly be realized in Nature.

\section*{Acknowledgements}
We would to thank Marina Cobal and Nicola Venturi for discussion of multi-photon LHC searches. DB thanks Andrea Romanino and Marco Serone for useful discussion.

\appendix

\section{The $L+N$ model}
\label{appendixA}

In this appendix we describe in detail the model with constituents fermions $L+N$ of $SU(N)$ \cite{Antipin:2015jia} that we use as a case study in Section~\ref{collider}.

The $O(p^2)$ low energy effective lagrangian, given in Eq. (\ref{lagrangianN}), reads
\begin{eqnarray}
\mathcal{L}&=&\frac{f^2}{4}{\rm Tr} [D_\mu U D^\mu U^\dagger]+  (g_\rho f^3 {\rm Tr}[MU^\dagger]+ h.c) + \frac {3 g_2^2 g_\rho^2 f^4 } {2(4\pi)^2} \sum_{i=1..3} {\rm Tr}[U T^i U^\dagger T^i]
\label{lagrangianN2}
\end{eqnarray}
where, 
\be
M = \left( \begin{array}{ccc}
m_L & 0 & y h^+ \\
0 & m_L & y h^0 \\
\tilde{y}  h^{-} & \tilde{y}  h^{0\dagger}  & m_N \\  
\end{array} \right)\, \qquad \quad  {\rm and} \qquad \quad U\equiv e^{i\sqrt{2}\Pi/f}
\label{massmatrix}
\ee
with
\begin{equation}
\Pi = \left( \begin{array}{ccc}
\pi_3^0/\sqrt{2}+\eta/\sqrt{6}& \pi_3^+ & K^{+} \\
\pi_3^{-} & -\pi_3^0/\sqrt{2}+\eta/\sqrt{6} & K^{0} \\
K^{-} & \bar{K}^{0} & -2\eta/\sqrt{6} \end{array} \right)  \quad  {\rm and} \quad H=(h^+,h^0)^T, \quad K=(K^+,K^0)^T.
\label{QCDpions}
\end{equation}
The covariant derivative takes the form $D_\mu U=\partial_\mu U -i  A_\mu U+i U  A_\mu$ where $A_\mu$ are the SM gauge fields. 
We assume that the $\theta$  angle in the dark sector  has been rotated to the mass matrix so that couplings and masses are in general complex. 

Expanding the effective lagrangian around the $\Pi=0$ one finds several contributions to the scalar potential. There are terms that are CP conserving and terms that are CP violating, we present the full expression in the following form
\be
V = - g_\rho f^3 {\rm Re}[4 m_L+ 2 m_N] + V_{y,H} + V_{CP}^{\pi} + V_{\slashed{CP}}^{\pi}
\ee
The first term contributes to the cosmological constant term and plays no role in our analysis, $V_{y,H}$ contains the potential terms that involve one insertion of the Higgs field (and one insertion of Yukawa coupling), while $V^\pi$ contains only terms with NGBs and we differentiate between contributions that respect or violate CP. Explicitly one finds
\begin{eqnarray}
V_{y,H}&=& i\sqrt{2} g_\rho f^2 y_- K^{\dagger} H ( 1 - \frac{\pi_3^2+\eta^2 + 2|K|^2}{6f^2}) +\frac{g_\rho}{\sqrt{2}} y_+f \left(K^{\dagger} \sigma^a\pi^{a}_3 -\frac{\eta K^{\dagger}}{\sqrt{3}} \right) H + h.c.\,\\
V_{CP}^\pi &=& m_K^2 |K|^2 + \frac{1}{2}m_\pi^2 \pi^2 + \frac{1}{2}m_\eta^2 \eta^2 + \lambda_K |K|^4 + \lambda_\pi \pi^4 + \lambda_\eta \eta^4 +\lambda_{\eta K} \eta^2 |K|^2\nonumber\\
& +&\lambda_{\eta \pi} \eta^2 \pi^2 +\lambda_{\pi K} \pi^2 |K|^2 +\lambda_4 \eta K^\dag \pi K\\
V_{\slashed{CP}}^\pi &=& t \eta + a_\eta \eta^3 + a_{\eta \pi}\eta \pi^2 + a_{\eta K} \eta |K|^2 + a_{\pi K} K^\dag \pi K
\end{eqnarray}
where  the mass terms are given by
\begin{eqnarray}
\label{pionmasses}
m^2_{\pi_3}&\approx& \frac{6 g_2^2 g_\rho^2 }{(4\pi)^2}f^2+ 4 {\rm Re}[m_L]g_{\rho} f \\
m^2_{K}&\approx&  \frac{9 g_2^2 g_\rho^2}{4(4\pi)^2}f^2+ 2{\rm Re}[m_L+m_N]g_{\rho} f\\
m_{\eta}^2&\approx& \frac{4}{3}{\rm Re}[m_L+2m_N]g_{\rho} f \ . \nonumber
\end{eqnarray}
The quartic couplings are  explicitly given by
\be
\lambda_K = - \frac{g_\rho}{3f}\re[m_L+m_N],\quad \lambda_\pi = - \frac{g_\rho}{6f}\re[m_L],\quad \lambda_\eta = - \frac{g_\rho}{54f}\re[m_L+8m_N]
\ee
\be
\lambda_{\eta K} = - \frac{g_\rho}{6f}\re[m_L+3m_N],\quad \lambda_{\eta\pi} = - \frac{g_\rho}{3f}\re[m_L],\quad \lambda_{\pi K} = - \frac{g_\rho}{6f}\re[3m_L+m_N],\quad \lambda_{4} =  \frac{g_\rho}{6\sqrt{3}f}\re[m_L-m_N]
\ee
while the CP violating parameters are
\be
t =-4 \frac{g_\rho f^2}{\sqrt{3}} \im[m_L-m_N]\,,\quad a_{\eta} =\frac{2 g_\rho}{9 \sqrt{3}} \im[m_L-4m_N]\,\quad a_{\eta K}=-\frac{2g_\rho}{\sqrt{3}}\im[m_N]\,\quad \ee
\be
a_{\eta \pi}=\frac{2g_\rho}{\sqrt{3}}\im[m_L]\,\quad a_{\pi K}=-\frac{2g_\rho}{3}\im[2m_L+m_N].
\ee
The singlet tadpole $t$ can be avoided rotating the phases so that
${\rm Im}(m_L)- {\rm Im}(m_N)=0$. Note, however, that for complex masses a CP violating coupling $\eta(\pi) KK$ (and $\eta^3$) is in general generated.

\section{Multiphoton backgrounds}
\label{appendixB}

In this Appendix we describe in details the background simulations that we have performed in Sec.~\ref{3gamma} in order to estimate the reach of our proposed search strategy for the $3\gamma$ final state.
As mentioned in the  text, the main SM backgrounds for this signature are 
real $3\gamma + nj$, as well as $2\gamma + nj$ processes, where additional photons arise from initial/final state radiation effects (ISR/FSR) and jet mis-identified as photons. 
 
We compute these backgrounds
parametrizing the probability to misidentify a jet as a photon as ${\cal P}_{j\to \gamma}=0.0093 e^{-0.036 p_T^j/{\rm GeV}}$ for $p_T>28\;$GeV and  $9.5\times 10^{-4} + 1.5\times 10^{-4}\times p_T/{\rm GeV}$ for $15\;$GeV$<p_T<28\;{\rm GeV}$ and rescaling the photon $p_T$  by a factor 0.75~\cite{ATL-PHYS-PUB-2013-009,Jaeckel:2015jla}.
 In order to validate our simulation we compare it  with the background yields in the $3\gamma$ signal region as defined in the ATLAS analysis~\cite{Aad:2015bua} and with the ones obtained by Ref.~\cite{Freitas:2010ht}, where a preliminary analysis of the LHC reach on the $3\gamma W$ final state was performed. As regarding the ATLAS analysis, the comparison is done with respect to the number of events in the signal region defined by the presence of 3 isolated photons with $p_T^{\gamma_{1,2,3}}>22,22,17\;$GeV, while for~\cite{Aad:2015bua} we compare the yields obtained at generator level requiring $p_T^\gamma>40
\;$ GeV and, at reconstructed level,  the presence of 3 isolated photons with the same transverse momenta requirements.
 The numerical comparison are reported in Tab.~\ref{tab:bkg-comparison}.
 \begin{table}[h!]
 \begin{center}
  \begin{tabular}{ c || c | c }
  \multicolumn{3}{c}{\textbf{Comparison with ATLAS~\cite{Aad:2015bua}}}\\
  \hline
    Process 			& \cite{Aad:2015bua} [fb] & Our [fb] \\     
    \hline
    $3\gamma$	& 16.7						& 18.4  \\
    \hline
    $2\gamma$ j	& 17.2						&  83.4  \\    
    \hline
  \end{tabular}
  \\
  \vskip 25pt
    \begin{tabular}{ c || c | c || c | c}
  \multicolumn{5}{c}{\textbf{Comparison with~\cite{Freitas:2010ht}}}\\   
  \hline 
    Process 			& \cite{Freitas:2010ht}  Gen. [fb] & Our  Gen. [fb] & \cite{Freitas:2010ht}  Reco. [fb] & Our  Reco. [fb] \\     
    \hline
  $3\gamma + \{0,1,2\}j$ 	 & 2.5  								& 3.7  & 2.0  								& 1.6   \\
    \hline
  $2\gamma + \{0,1,2\}j$ 	 & 7.2$\times 10^3$  								& 9.7$\times 10^3$  	&  		5.9						&  4.7 \\  
    \hline
  \end{tabular}
\end{center}
\caption{\em Comparison with the ATLAS 8 TeV analysis~\cite{Aad:2015bua} and with the 7 TeV estimation of~\cite{Freitas:2010ht}. The ATLAS analysis signal region is defined with $p_T^{\gamma_{1,2,3}}>22,22,17\;$ GeV while the rates at generator level of~\cite{Freitas:2010ht} are computed with $p_T^\gamma>40\;$ GeV. The same cut is applied at reconstructed level, where three isolated photons are required.
}
\label{tab:bkg-comparison}
\end{table}
 Note that the ATLAS analysis splits the main backgrounds in $2\gamma$, $3\gamma$ and $2\gamma j$ processes, where from the $2\gamma$ rate events, where one photon arises from a mis-identified jet, have been removed to avoid double counting with the $2\gamma j$ sample. All together we thus cross-checked our simulation only against the $3\gamma$ and $2\gamma j$ estimation from ATLAS. After having applied NLO QCD $\kappa$-factors, estimated by running {\tt
MadGraph5\_aMC@NLO} at LO and NLO,  
we found that  we reproduce the real background within $\sim 10\%$ of the ATLAS estimation while our $2\gamma j$ rate is found to be a factor $\sim\;$5 larger. 
The $2\gamma j$ background  is however estimated experimentally via data driven techniques, which are hard to be reproduced with a fast detector simulation, and strongly depend on the parametrization chosen for the $\gamma\to j$ mis-identification rate.
As regarding the comparison with~\cite{Freitas:2010ht} we find a $\sim 30\%$ agreement for all the background estimations. In this case, in doing the comparison,
our rates have been multiplied by a flat $\kappa$-factor of 2, as done in~\cite{Freitas:2010ht}. All together we find a quite good agreement with existing experimental and theoretical results, thus making the analysis performed in Sec.~\ref{3gamma} solid.


\begin{thebibliography}{nn}

\bibitem{Kilic:2009mi}
  C.~Kilic, T.~Okui and R.~Sundrum,
  ``Vectorlike Confinement at the LHC,''
  JHEP {\bf 1002} (2010) 018
  doi:10.1007/JHEP02(2010)018
  [arXiv:0906.0577 [hep-ph]].
  
\bibitem{Antipin:2015jia} 
  O.~Antipin and M.~Redi,
  ``The Half-composite Two Higgs Doublet Model and the Relaxion,''
  JHEP {\bf 1512}, 031 (2015)
  doi:10.1007/JHEP12(2015)031
  [arXiv:1508.01112 [hep-ph]].
  
  \bibitem{Agugliaro:2016clv} 
  A.~Agugliaro, O.~Antipin, D.~Becciolini, S.~De Curtis and M.~Redi,
  ``UV complete composite Higgs models,''
  Phys.\ Rev.\ D {\bf 95}, no. 3, 035019 (2017)
  doi:10.1103/PhysRevD.95.035019
  [arXiv:1609.07122 [hep-ph]].   
  
  \bibitem{Panico:2015jxa} 
  G.~Panico and A.~Wulzer,
  ``The Composite Nambu-Goldstone Higgs,''
  Lect.\ Notes Phys.\  {\bf 913}, pp.1 (2016)
  doi:10.1007/978-3-319-22617-0
  [arXiv:1506.01961 [hep-ph]].  
  
   \bibitem{ACDM} 
O.~Antipin, M.~Redi, A.~Strumia and E.~Vigiani,
  ``Accidental Composite Dark Matter,''
  arXiv:1503.08749 [hep-ph].
  
  \bibitem{Batell:2017kho} 
  B.~Batell, M.~A.~Fedderke and L.~T.~Wang,
 JHEP {\bf 1712}, 139 (2017)
  doi:10.1007/JHEP12(2017)139
  [arXiv:1705.09666 [hep-ph]].

 

  
\bibitem{Graham:2015cka} 
  P.~W.~Graham, D.~E.~Kaplan and S.~Rajendran,
  ``Cosmological Relaxation of the Electroweak Scale,''
  Phys.\ Rev.\ Lett.\  {\bf 115}, no. 22, 221801 (2015)
  doi:10.1103/PhysRevLett.115.221801
  [arXiv:1504.07551 [hep-ph]].  
  

\bibitem{Mitridate:2017oky} 
  A.~Mitridate, M.~Redi, J.~Smirnov and A.~Strumia,
  ``Dark Matter as a weakly coupled Dark Baryon,''
  JHEP {\bf 1710}, 210 (2017)
  [arXiv:1707.05380 [hep-ph]]. 
  
\bibitem{Elor:2018xku} 
  G.~Elor, H.~Liu, T.~R.~Slatyer and Y.~Soreq,
  ``Complementarity for Dark Sector Bound States,''
  arXiv:1801.07723 [hep-ph].    
  
\bibitem{Mrazek:2011iu} 
  J.~Mrazek, A.~Pomarol, R.~Rattazzi, M.~Redi, J.~Serra and A.~Wulzer,
  ``The Other Natural Two Higgs Doublet Model,''
  Nucl.\ Phys.\ B {\bf 853}, 1 (2011)
  doi:10.1016/j.nuclphysb.2011.07.008
  [arXiv:1105.5403 [hep-ph]].  
  
\bibitem{Giudice:2007fh}
  G.~F.~Giudice, C.~Grojean, A.~Pomarol and R.~Rattazzi,
  JHEP {\bf 0706} (2007) 045
  doi:10.1088/1126-6708/2007/06/045
  [hep-ph/0703164].
  
  \bibitem{Branco:2011iw}
  G.~C.~Branco, P.~M.~Ferreira, L.~Lavoura, M.~N.~Rebelo, M.~Sher and J.~P.~Silva,
  ``Theory and phenomenology of two-Higgs-doublet models,''
  Phys.\ Rept.\  {\bf 516} (2012) 1
  doi:10.1016/j.physrep.2012.02.002
  [arXiv:1106.0034 [hep-ph]].  
  
 \bibitem{Barbieri:2004qk} 
  R.~Barbieri, A.~Pomarol, R.~Rattazzi and A.~Strumia,
  ``Electroweak symmetry breaking after LEP-1 and LEP-2,''
  Nucl.\ Phys.\ B {\bf 703}, 127 (2004)
  doi:10.1016/j.nuclphysb.2004.10.014
  [hep-ph/0405040].  
  
  \bibitem{Farina:2016rws}
    M.~Farina, G.~Panico, D.~Pappadopulo, J.~T.~Ruderman, R.~Torre and A.~Wulzer,
    Phys.\ Lett.\ B {\bf 772} (2017) 210
    doi:10.1016/j.physletb.2017.06.043
    [arXiv:1609.08157 [hep-ph]].
	
	\bibitem{Baak:2012kk}
	  M.~Baak {\it et al.},
	  Eur.\ Phys.\ J.\ C {\bf 72} (2012) 2205
	  doi:10.1140/epjc/s10052-012-2205-9
	  [arXiv:1209.2716 [hep-ph]].
  
  \bibitem{MFV}
  G.~D'Ambrosio, G.~F.~Giudice, G.~Isidori and A.~Strumia,
  ``Minimal flavor violation: An Effective field theory approach,''
  Nucl.\ Phys.\ B {\bf 645} (2002) 155
  doi:10.1016/S0550-3213(02)00836-2
  [hep-ph/0207036].
  
  \bibitem{Enomoto:2015wbn} 
  T.~Enomoto and R.~Watanabe,
  ``Flavor constraints on the Two Higgs Doublet Models of Z$_{2}$ symmetric and aligned types,''
  JHEP {\bf 1605}, 002 (2016)
  doi:10.1007/JHEP05(2016)002
  [arXiv:1511.05066 [hep-ph]].
  
 \bibitem{Meroni}
  T.~Alanne, D.~Buarque Franzosi, M.~T.~Frandsen, M.~L.~A.~Kristensen, A.~Meroni and M.~Rosenlyst,
  ``Partially composite Higgs models: Phenomenology and RG analysis,''
  JHEP {\bf 1801} (2018) 051
  doi:10.1007/JHEP01(2018)051
  [arXiv:1711.10410 [hep-ph]].  

  
\bibitem{Baron:2013eja} 
  J.~Baron {\it et al.} [ACME Collaboration],
  ``Order of Magnitude Smaller Limit on the Electric Dipole Moment of the Electron,''
  Science {\bf 343}, 269 (2014)
  doi:10.1126/science.1248213
  [arXiv:1310.7534 [physics.atom-ph]].
    
  \bibitem{Kilic:2010et} 
    C.~Kilic and T.~Okui,
    ``The LHC Phenomenology of Vectorlike Confinement,''
    JHEP {\bf 1004}, 128 (2010)
    [arXiv:1001.4526 [hep-ph]].
	
  
    \bibitem{Mariotti:2017vtv} 
    A.~Mariotti, D.~Redigolo, F.~Sala and K.~Tobioka,
    ``New LHC bound on low-mass diphoton resonances,''
    arXiv:1710.01743 [hep-ph].
	
	\bibitem{Cirelli:2005uq} 
	  M.~Cirelli, N.~Fornengo and A.~Strumia,
	  Nucl.\ Phys.\ B {\bf 753}, 178 (2006)
	  doi:10.1016/j.nuclphysb.2006.07.012
	  [hep-ph/0512090].
	
	
  \bibitem{CMSZZ} 
  A.~M.~Sirunyan {\it et al.} [CMS Collaboration],
  ``Search for a new scalar resonance decaying to a pair of Z bosons in proton-proton collisions at $\sqrt{s} =$ 13 TeV,''
  arXiv:1804.01939 [hep-ex]. 
  
  \bibitem{Aaboud:2017rel}
    M.~Aaboud {\it et al.} [ATLAS Collaboration],
    Eur.\ Phys.\ J.\ C {\bf 78} (2018) no.4,  293
    doi:10.1140/epjc/s10052-018-5686-3
    [arXiv:1712.06386 [hep-ex]].
	
	\bibitem{Aad:2015bua}
	  G.~Aad {\it et al.} [ATLAS Collaboration],
	  Eur.\ Phys.\ J.\ C {\bf 76} (2016) no.4,  210
	  doi:10.1140/epjc/s10052-016-4034-8
	  [arXiv:1509.05051 [hep-ex]].
	  
\bibitem{Alloul:2013bka}
  A.~Alloul, N.~D.~Christensen, C.~Degrande, C.~Duhr and B.~Fuks,
  Comput.\ Phys.\ Commun.\  {\bf 185} (2014) 2250
  doi:10.1016/j.cpc.2014.04.012
  [arXiv:1310.1921 [hep-ph]].
  
\bibitem{Degrande:2011ua}
C.~Degrande, C.~Duhr, B.~Fuks, D.~Grellscheid, O.~Mattelaer and T.~Reiter,
    Comput.\ Phys.\ Commun.\  {\bf 183} (2012) 1201
    doi:10.1016/j.cpc.2012.01.022
    [arXiv:1108.2040 [hep-ph]].
	
\bibitem{Alwall:2014hca}
  J.~Alwall {\it et al.},
  JHEP {\bf 1407} (2014) 079
  doi:10.1007/JHEP07(2014)079
  [arXiv:1405.0301 [hep-ph]].
  
\bibitem{Sjostrand:2007gs}
  T.~Sjostrand, S.~Mrenna and P.~Z.~Skands,
  Comput.\ Phys.\ Commun.\  {\bf 178} (2008) 852
  doi:10.1016/j.cpc.2008.01.036
  [arXiv:0710.3820 [hep-ph]].
  
	\bibitem{deFavereau:2013fsa}
	  J.~de Favereau {\it et al.} [DELPHES 3 Collaboration],
	  JHEP {\bf 1402} (2014) 057
	  doi:10.1007/JHEP02(2014)057
	  [arXiv:1307.6346 [hep-ex]].
	  
	 
  
\bibitem{Aaboud:2017buh}
  M.~Aaboud {\it et al.} [ATLAS Collaboration],
  JHEP {\bf 1710} (2017) 182
  doi:10.1007/JHEP10(2017)182
  [arXiv:1707.02424 [hep-ex]].
  
 \bibitem{CMS:2016ybj}
   CMS Collaboration [CMS Collaboration],
   CMS-PAS-EXO-16-036.
	
  
    \bibitem{Gripaios:2009pe}
  B.~Gripaios, A.~Pomarol, F.~Riva and J.~Serra,
  JHEP {\bf 0904} (2009) 070
  doi:10.1088/1126-6708/2009/04/070
  [arXiv:0902.1483 [hep-ph]].
  
\bibitem{Aad:2014yja}
  G.~Aad {\it et al.} [ATLAS Collaboration],
  Phys.\ Rev.\ Lett.\  {\bf 114} (2015) no.8,  081802
  doi:10.1103/PhysRevLett.114.081802
  [arXiv:1406.5053 [hep-ex]].
  
\bibitem{Khachatryan:2016sey}
  V.~Khachatryan {\it et al.} [CMS Collaboration],
  Phys.\ Rev.\ D {\bf 94} (2016) no.5,  052012
  doi:10.1103/PhysRevD.94.052012
  [arXiv:1603.06896 [hep-ex]].
 
\bibitem{TheATLAScollaboration:2016ibb}
  The ATLAS collaboration,
  ATLAS-CONF-2016-004.
  
\bibitem{CMS:2017ihs}
  CMS Collaboration [CMS Collaboration],
  CMS-PAS-HIG-17-008.
  

  


  
\bibitem{Aaboud:2017sjh}
  M.~Aaboud {\it et al.} [ATLAS Collaboration],
  JHEP {\bf 1801} (2018) 055
  doi:10.1007/JHEP01(2018)055
  [arXiv:1709.07242 [hep-ex]].
  
\bibitem{Khachatryan:2015sma}
  V.~Khachatryan {\it et al.} [CMS Collaboration],
  Phys.\ Rev.\ D {\bf 93} (2016) no.1,  012001
  doi:10.1103/PhysRevD.93.012001
  [arXiv:1506.03062 [hep-ex]].  
  
  \bibitem{Bai:2010qg} 
    Y.~Bai and R.~J.~Hill,
    Phys.\ Rev.\ D {\bf 82}, 111701 (2010)
    doi:10.1103/PhysRevD.82.111701
    [arXiv:1005.0008 [hep-ph]].
    
 \bibitem{Aaboud:2017qph}
   M.~Aaboud {\it et al.} [ATLAS Collaboration],
   arXiv:1710.09748 [hep-ex].

  
\bibitem{ATL-PHYS-PUB-2013-009}
  G.~Aad {\it et al.} [ATLAS Collaboration],
  ATL-PHYS-PUB-2013-009
http://cds.cern.ch/record/1604420	

\bibitem{Jaeckel:2015jla}
  J.~Jaeckel and M.~Spannowsky,
  Phys.\ Lett.\ B {\bf 753} (2016) 482
  doi:10.1016/j.physletb.2015.12.037
  [arXiv:1509.00476 [hep-ph]].

\bibitem{Freitas:2010ht}
  A.~Freitas and P.~Schwaller,
  JHEP {\bf 1101} (2011) 022
  doi:10.1007/JHEP01(2011)022
  [arXiv:1010.2528 [hep-ph]].
	  
  



  

  
  
\end{thebibliography}
\end{document}